\documentclass[fleqn,usenatbib]{mnras}

\usepackage[T1]{fontenc}

\DeclareRobustCommand{\VAN}[3]{#2}
\let\VANthebibliography\thebibliography
\def\thebibliography{\DeclareRobustCommand{\VAN}[3]{##3}\VANthebibliography}

\usepackage{graphicx}	
\usepackage{amsmath}	
\usepackage{amssymb}	
\usepackage{xspace}
\usepackage{braket}
\usepackage{booktabs}

\RequirePackage{lineno}
\usepackage[utf8]{inputenc}
\usepackage{ae,aecompl}
\usepackage{comment}
\usepackage{natbib}
\usepackage{float}
\usepackage{graphicx}
\usepackage{subcaption}
\usepackage{amssymb}
\usepackage{amsmath}
\usepackage{rotating,times,pictex,graphicx,latexsym}
\usepackage{color}
\usepackage{amsmath}
\usepackage{threeparttable}
\usepackage{lipsum}

\usepackage[all]{hypcap}
\usepackage[title,titletoc]{appendix}
\usepackage[]{hyperref}
\PassOptionsToPackage{pdfpagelabels=false}{hyperref}
\usepackage[T1]{fontenc}
\usepackage{newtxtext,newtxmath}

\newcommand{\Ms}{M\ensuremath{_{\odot}}}

\newcommand{\muG}{$\mu$G}

\newcommand{\beq}{\begin{equation}}
\newcommand{\eeq}{\end{equation}}

\newcommand{\mum}{\,\ensuremath{\mu}m\xspace}

\newcommand{\mach}{\hbox{$\mathcal{M}_A$}}
\newcommand{\arcs}{\hbox{$^{\prime\prime}$}}

\newcommand{\arcm}{\mbox{$^{\prime}$}}

\newcommand{\hii}{\mbox{H~{\sc {ii}}~}}

\newcommand{\degree}{\mbox{$^{\circ}$}}

\newcommand{\alf}{Alfv$\acute{\text{e}}$n~} 
\newcommand{\mjb}{\hbox{mJy beam$^{-1}$}}

\newcommand{\kms}{\hbox{km~s$^{-1}$}}
\newcommand{\cms}{\hbox{cm$^{-2}$~}}

\newcommand{\cmq}{\hbox{cm$^{-3}$}}

\newcommand{\thco}{\hbox{$^{13}$CO}~}
\newcommand{\twco}{\hbox{$^{12}$CO}~}
\newcommand{\eico}{\hbox{C$^{18}$O}~}

\newcommand{\cloud}{{\rm G148.24+00.41}}

\title[Magnetic field at Hub of GMC G148.24+00.41]{
\center
Understanding the relative importance of magnetic field, gravity, and turbulence in star formation at the hub
of the giant molecular cloud G148.24+00.41}

\author[Rawat et al.]{
Vineet Rawat,$^{1,2}$\thanks{E-mail: vineet@prl.res.in}
M. R. Samal,$^{1}$
Chakali Eswaraiah,$^{3}$
Jia-Wei Wang,$^{4}$
Davide Elia,$^{5}$\newauthor
Sandhyarani Panigrahy,$^{3}$
A. Zavagno,$^{6,7}$
R. K. Yadav,$^{8}$
D. L. Walker,$^{9}$
J. Jose,$^{3}$
D.K. Ojha,$^{10}$\newauthor
C.P. Zhang,$^{11,12}$
S. Dutta,$^{4}$
\\
$^{1}$Physical Research Laboratory, Navrangpura, Ahmedabad, Gujarat 380009, India\\
$^{2}$Indian Institute of Technology Gandhinagar Palaj, Gandhinagar 382355, India\\
$^{3}$Indian Institute of Science Education and Research (IISER) Tirupati, Rami Reddy Nagar, Karakambadi Road, Mangalam (P.O.), Tirupati 517 507, India\\
$^{4}$Academia Sinica Institute of Astronomy and Astrophysics, No.1, Sec. 4, Roosevelt Road, Taipei 10617, Taiwan\\
$^{5}$Istituto di Astrofisica e Planetologia Spaziali, INAF, Via Fosso del Cavaliere 100, I-00133 Roma, Italy\\
$^{6}$Aix-Marseille Universite, CNRS, CNES, LAM, 38 rue F. Joliot Curie, F-13388 Marseille Cedex 13, France\\
$^{7}$Institut Universitaire de France, Paris, 1 rue Descartes, F-75231 Paris Cedex 05, France\\
$^{8}$National Astronomical Research Institute of Thailand (NARIT), Sirindhorn AstroPark, 260 Moo 4, T. Donkaew, A. Maerim, Chiangmai 50180, Thailand\\
$^{9}$Jodrell Bank Centre for Astrophysics, Department of Physics and Astronomy, University of Manchester, Oxford Road, Manchester M13 9PL, UK\\
$^{10}$Department  of  Astronomy  and  Astrophysics,  Tata  Institute  of  Fundamental  Research,  Mumbai  400005, India\\
$^{11}$National Astronomical Observatories, Chinese Academy of Sciences, Beijing 100101, People’s Republic of China\\
$^{12}$Guizhou Radio Astronomical Observatory, Guizhou University, Guiyang 550000, People’s Republic of China\\
}
\begin{document}
\label{firstpage}
\pagerange{\pageref{firstpage}--\pageref{lastpage}}
\maketitle
\begin{abstract}
The relative importance of magnetic fields, turbulence, and gravity in the early phases of
star formation is still not well understood. We report the first high-resolution dust polarization observations at 850 $\mu$m around the most massive clump, located at the hub of the Giant Molecular Cloud \cloud, using SCUBA-2/POL-2 at the James Clerk Maxwell Telescope. We find that the degree of polarization decreases steadily towards the denser portion of the cloud. Comparing the intensity gradients and local gravity with the magnetic field orientations, we find that local gravity plays a dominant role in driving the gas collapse as the magnetic field orientations and gravity vectors seem to point towards the dense clumps. We also find evidence of U-shaped magnetic field morphology towards a small-scale elongated structure associated with the central clump, hinting at converging accretion flows towards the clump. Our observation has resolved the massive clump into multiple substructures. We study the magnetic field properties of two regions, central clump (CC) and northeastern elongated structure (NES). Using the modified Davis-Chandrasekhar-Fermi method, we determine that the magnetic field strengths of CC and NES are $\sim$24.0 $\pm$ 6.0 \muG~and 20.0 $\pm$ 5.0 $\mu$G, respectively. The mass-to-flux ratios are found to be magnetically transcritical/supercritical, while the \alf Mach number indicates a trans-Alfv$\acute{\text{e}}$nic state in both regions. These results, along with Virial analysis, suggest that at the hub of \cloud, gravitational energy has an edge over magnetic and kinetic energies.

\end{abstract}




\section{Introduction}
\label{int}
Molecular clouds, the enigmatic birthplaces of stars, are now well established to consist of filamentary structures, revealed by $\it{Herchel}$ observations \citep{Molinari_2010, Andre_2013, Zavagno_2023}. Within these immense structures of cold gas and dust, the gravity, turbulence, and magnetic fields dictate the process of star formation at different scales, from large scale (cloud and filaments) to small scale (clumps and dense cores) \citep{Klessen_2000, Balles_2007, Federrath_2015, Tang_2019, Wang_2020b, Pattle_2022}. However, their relative role during the different stages of cloud evolution is still unclear and a topic of debate \citep{Li_2014}. 

The magnetic field exists throughout star-forming molecular clouds across various scales \citep[see the review article by][]{Pattle_2022} and plays a crucial role in the formation of molecular clouds and filamentary structures \citep{soler_2013, Hennebelle_2019}. Numerical simulations show that the strong magnetic fields play an important role in the magnetically channelled gravitational collapse of clouds \citep{Nakamura_2008, Gomez_2018}, and can channel turbulent flows along the filaments \citep{Li_2008, soler_2013, Zamora_2017}, guide the accreting matter \citep{Seif_2015, Shimajiri_2019}, and dynamically influence the formation of cores along the dense ridges of the filamentary clouds \citep[e.g.][]{koch_2014, Zhang_2014, cox_2016, Pattle_2017, Soam_2018, Liu_2019, Eswaraiahetal2021}. At parsec (or few parsec) scale, the magnetic field typically shows an ordered structure, mostly aligned with the long axis of the low-density elongated gas structures such as striations, while in high-density filaments, the magnetic field lines are preferentially perpendicular to the long axes of filaments \citep[e.g.][]{cox_2016, Planck_2016, soler_2017, ward_thomp_2017, Tang_2019, Soam_2019, Doi_2020}. At sub-parsec scales, the magnetic field can be very complex depending upon the turbulent nature of the magnetic field and stellar feedback \citep[e.g.][]{Hull_2017, Eswaraiahetal2020,  Eswaraiahetal2021}. Dust polarization studies at small scales have revealed a variety of magnetic field morphologies in dense clumps and cores, and the results suggest that the magnetic field is scale-dependent and varies with the environment \citep[e.g.][]{Girart_2013, Hull_2017, ward_thomp_2017, Pattle_2018, Eswaraiahetal2020, Eswaraiahetal2021}.

Moreover, it is not clear whether it is turbulence or magnetic field along with gravity that dominates the star formation process. The role of gravity has long been recognised as the primary factor driving the collapse of dense regions and initiating the birth of protostellar cores. On the other hand, turbulence, the chaotic and ubiquitous motion of gas within these massive clouds, influences the fragmentation of the collapsing gas. However, the role of magnetic field, in comparison to turbulence and gravity, is relatively less understood at various stages of star formation.

The ``strong magnetic field'' theory of star formation stresses the importance of the magnetic field in the formation and evolution of clouds and subsequent structures \citep{Mouschovias_2006, Tan_2013, Hennebelle_2018}. 
Conversely, the ``weak magnetic field'' theory suggests that turbulent flows control the formation and evolution of clouds and cores, and create the compressed regions where stars form \citep{Padoan_2002, Maclow_klessen_2004, Fed_kle_2012}. 
Observationally also, in some high-mass star-forming regions, it has been found that the turbulent energy is more dominant or comparable to magnetic energy \citep[e.g.][]{Beuther_2010, Girart_2013, Beuther_2020, Wang_2020b}. In contrast, other studies found magnetic energy to be more dominant than turbulent energy \citep[e.g.][]{Girart_2009, Beuther_2018, Eswaraiahetal2020, Chung_2023}. 
Therefore, more observational evidence is required at the early stages of star formation to better constrain the theoretical models and relative roles of gravity, magnetic field, and turbulence in the star-forming regions.

Dust polarization observation is a key tool for tracing the plane-of-sky (POS) magnetic field geometry in star-forming regions. The polarization is caused by asymmetric dust grains that preferentially align their shorter axis along the magnetic field direction \citep{Laz_2007, Hoang_2008}. There are different mechanisms for dust grain alignment  \cite[see the review article by][]{And_2015}, but the most widely accepted one being is the radiative alignment torque (RAT) mechanism \citep{Laz_2007, Hoang_2014, And_2015}. 

Located at a distance of $\sim$3.4 kpc, \cloud~is a massive cloud having mass $\sim$10$^5$ \Ms~and dust temperature $\sim$14.5 K, and it resembles a hub-filament morphology \citep{Rawat_2023}. The cloud is still at the early stages of its evolution such that the stellar feedback mechanisms like stellar winds and expanding \hii regions are not yet significant \citep{Rawat_2023}, which makes G148.24+00.41 a potential candidate to study the early stages of the star and cluster formation process. Fig. \ref{fig:scan}a shows the \thco intensity map of \cloud, integrated in the velocity range of $-$37.0 \kms~to $-$30.0 \kms, where
one can see a bright spot in the centre of the map. This location corresponds to the most massive clump of the cloud, which is nomenclatured as C1 in \cite{Rawat_2024}. \cite{Rawat_2024}, using CO molecular data (spatial resolution $\sim$52\arcsec or 0.9 pc at distance $\sim$3.4 kpc, velocity resolution $\sim$0.17 \kms), also found filamentary features and seven massive clumps in \cloud. The C1 clump with a mass of $\sim$2100 \Ms~and an effective size of $\sim$1.8 pc based on \eico data, is found to be at the hub/central location of the cloud. The filamentary features attached to the C1 clump can also be seen in $\it{Herschel}$ 250 \mum image, shown in Fig. \ref{fig:scan}b \citep[adapted from][]{Rawat_2023}. The figure clearly reveals the hub-filament system morphology of the cloud. 
\cite{Rawat_2023} found that the clump hosts an infrared cluster, seen in $\it{Spitzer}$ bands, and hypothesised that the cluster might grow to a richer cluster by accumulating gas from the extended reservoir. Such hub filamentary systems with the clump being located at the nexus or junction of these filaments are of particular interest because these are the sites where cluster formation would take place, as advocated in simulations and observations \citep[e.g.][]{Naranjo_2012, Gomez_2014, Gomez_2018, sema2019, Kum_2020}. 
\\

The magnetohydrodynamic simulations suggest that the filamentary converging flows would impact the magnetic field morphology of the star-forming regions \citep{Gomez_2018}. In \cloud, \cite{Rawat_2024} found converging gas flows along the filaments towards the hub. So, it would be interesting to investigate the influence of gas flows on the B-field morphology in the hub of the cloud. In addition, it is equally important to examine the role of magnetic fields in the stability of such hub systems, as it is expected that gravity would play a dominant role in the onset of star formation within such systems. In this work, we investigate the morphology and strength of the magnetic field of the C1 clump of \cloud~for the first time, in order to understand the relative importance of the magnetic field in comparison to gravity and turbulence in the overall star formation process of the clump. This paper is organised as follows. In Section \ref{obs}, we describe the observations, data reduction, and other data sets used in this work. In Section \ref{results}, we present the analysis and results related to the magnetic field morphology, its relative orientation in comparison to intensity gradients and local gravity, and its strength. We also discuss the dust and gas properties of the studied region. In Section \ref{discussion}, we discuss the results and compare the strength of gravity, magnetic field, and turbulence. In Section \ref{summary}, we summarize the findings of this work. 



\section{Observations and Data}
\label{obs}
\subsection{Dust continuum polarization observations using JCMT SCUBA-2/POL-2}

We observed the C1 clump/hub region of \cloud~with SCUBA-2/POL-2 mounted on the James Clerk Maxwell Telescope (JCMT), a single-dish sub-millimeter telescope in Mauna Kea, Hawaii, USA. The POL-2 instrument is a linear polarimetry module \citep{Friberg_2016} for the Submillimetre Common User Bolometer Array-2 (SCUBA-2), a 10,000 bolometer camera on the JCMT \citep{Holland_2013}. The data was acquired between 2022 November 25 and 2023 January 03 (project code: M22BP055; PI: Vineet Rawat) in the band 2 weather conditions under an atmospheric optical depth at 225 GHz ($\tau_{225}$) of 0.04 to 0.06. The observations were taken in 10 sets with an integration time of 30 minutes each, resulting in a total integration time of around 5.5 hr. The POL-2 DAISY scan mode \citep{Holland_2013, Friberg_2016} was adopted, which generates a map of high signal-to-noise ratio (SNR) within a central region spanning a diameter of 3\arcm, and the noise level gradually increases towards the edges of the map. 
The region is observed in both the 450 and 850 \mum continuum polarizations simultaneously, with a resolution of 9\farcs6 and 14\farcs1, respectively. Due to the low sensitivity of the 450 \mum data, this paper presents the analyses and results based on only 850 \mum dust polarization data.\\ 

\begin{figure*}
    \centering
    \includegraphics[width=17cm]{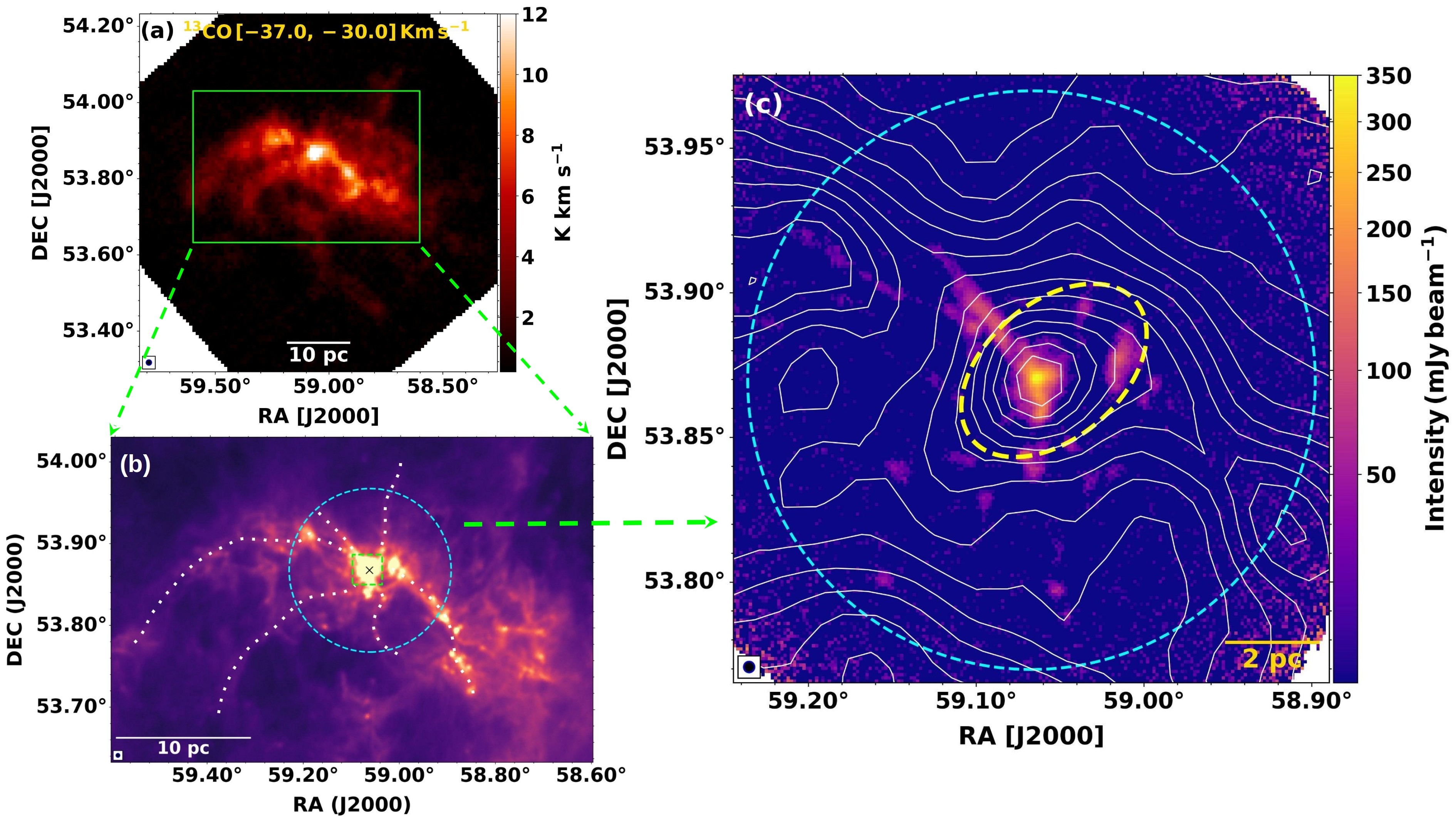}
    \caption{(a) \thco (J = 1$-$0) intensity map of \cloud, integrated in the velocity range $-$37.0 \kms to $-$30.0 \kms. (b) The central region (encompassed by the solid green box in panel-a) of \cloud~as seen in $\it{Herschel}$ 250 \mum band, showing the hub-filamentary morphology of the cloud. The figure is adapted from \protect\cite{Rawat_2023}, in which the blue circle shows the JCMT scanned region of diameter $\sim$12\arcm ($\sim$12 pc). The green dashed box marks the central area of the hub, where an infrared cluster is seen \citep{Rawat_2023}, and the cross sign indicates the position of a massive young stellar object. 
    (c) The 850 \mum Stokes I intensity map of the central region of \cloud~mapped by JCMT SCUBA-2/POL-2, along with the contours of \thco integrated intensity emission, drawn from 1.5 to 15 K \kms with a step size of $\sim$0.96 K \kms. The rms noise of the 4\arcsec pixel-size Stokes I map is around $\sim$5 mJy beam$^{-1}$. In panel-c, the yellow ellipse shows the position of the C1 clump, identified by \protect\cite{Rawat_2024} using \eico data (spatial resolution $\sim$52\arcsec). The beam sizes of the \thco integrated intensity map, $\it{Herschel}$ 250 \mum map, and JCMT 850 \mum map are $\sim$52\arcsec, 18\arcsec, and 14\arcsec, respectively, shown as a framed-blue dot at the bottom left of each panel.} 
    \label{fig:scan}
\end{figure*}

The data reduction was carried out using the $\it{pol2map}$\footnote{\href{http://starlink.eao.hawaii.edu/docs/sc22.htx/sc22.html}{http://starlink.eao.hawaii.edu/docs/sc22.htx/sc22.html}}script in the SMURF package \citep{Chapin_2013} of Starlink \citep{Currie_2014}. The POL-2 is characterized by linear polarization that produces Stokes I, Q, and U vector maps. 
The $\it{Skyloop}$ mode was utilised to minimise the uncertainty associated with map creation, while the MAPVARS mode was enabled to asses the total uncertainty from the standard deviation among individual observations. The details of the data reduction process of POL-2 can be found in  \cite{Pattle_2017} and \cite{Wang_2019}.
Finally, the I, Q, and U maps, along with their variance maps, are used to create a debiased polarization vector catalogue. The catalogue consists of total intensity (I), stokes vectors (Q and U), polarization intensity (PI), polarization fraction (P), polarization angle ($\theta_P$), and their associated uncertainties ($\delta$I, $\delta$Q, $\delta$U, $\delta$PI, $\delta$P, and $\delta$$\theta_P$, respectively).\\

The I, Q, and U maps are produced with 4\arcs pixel size, while the polarization catalogue is binned to 12\arcs, for better sensitivity. A flux calibration factor of 668.25~Jy beam$^{-1}$ pW$^{-1}$ is used for 850 \mum Stokes I, Q, and U map to convert them from pW to mJy/beam and to account for the flux-loss due to POL-2 insertion into the telescope. This calibration factor comprises 495 Jy/beam/pW for reductions using 4\arcs pixels of SCUBA-2, multiplied by the standard 1.35 factor for POL-2 losses \citep{Mairs_2021}.\\ 

The polarised intensity is defined to be positive, so the uncertainties of the Q and U Stokes vector would bias the polarised intensities towards larger values \citep{Vaill_2006, Kwon_2018}. The debiased polarization intensity and its uncertainty are calculated as 

\begin{equation}
    PI = \sqrt{Q^2 + U^2 - 0.5(\delta Q^2 + \delta U^2)}
\end{equation} and

\begin{equation*}
    \delta PI = \sqrt{\frac{(Q^2 \delta Q^2 + U^2 \delta U^2)}{(Q^2 + U^2)}},
\end{equation*} respectively. The debiased polarization fraction and its uncertainty are then calculated as

\begin{equation}
    P = \frac{PI}{I}
\end{equation} and

\begin{equation*}
    \delta P = \sqrt{\frac{\delta PI^2}{I^2} + \frac{\delta I^2 (Q^2 + U^2)}{I^4}},
\end{equation*} respectively. The polarization angle and its uncertainty are calculated as 

\begin{equation}
   \theta_P  = \frac{1}{2} \tan^{-1}{\left(\frac{U}{Q}\right)}
\end{equation} and

\begin{equation*}
    \delta_{\theta_P} = \frac{1}{2} \sqrt{\frac{(U^2 \delta Q^2 + Q^2 \delta U^2)}{(Q^2 + U^2)^2}},
\end{equation*} respectively. The polarization angle increases from the north toward the east, following the IAU convention. The mean rms noises in the Stokes I, Q, U, and PI measurements with 12$\arcsec$ bin size are 1.4, 1.1, 1.1, and 1.1 \mjb, respectively. Following the standard convention, for magnetic field, hereafter, B-field orientations, the polarization angles are rotated by 90 degrees. 

\subsection{Molecular line data from PMO}
We used the \thco (J = 1--0) and \eico (J = 1--0) molecular line data, which were taken as a part of the Milky Way Imaging Scroll Painting survey \citep[MWISP;][]{Su_2019} using the 13.7-m radio telescope at Purple Mount Observatory (PMO). The observation of CO isotopologues was taken simultaneously using a 3 $\times$ 3 beam sideband-separating Superconducting Spectroscopic Array Receiver (SSAR) system \citep{shan_2012} and using the position-switch on-the-ﬂy mode. The spatial resolution (Half Power Beam Width; HPBW) of \thco and \eico is around $\sim$52\arcsec ($\sim$0.9 pc at 3.4 kpc). And the spectral resolution of \thco and \eico is $\sim$0.17 \kms with a sensitivity of $\sim$0.3 K \citep[for details, see][]{Su_2019}.

\section{Analyses and Results}
\label{results}
Fig. \ref{fig:scan}c shows the Stokes I map of the region, where the location of the C1 clump is also shown. 
From the figure, it can be seen that the 850 \mum JCMT data (beam size $\sim$14\arcsec) has resolved multiple sub-structures in the central region of the cloud. 
We found sub-structures like a central clump, a clump located on the western side, and a prominent elongated structure on the northeastern side of the central clump. In addition to these prominent sub-structures, a few compact structures are also visible in the image. 

\subsection{B-field morphology}

\begin{figure*}
    \centering
    \includegraphics[width=8.5cm]{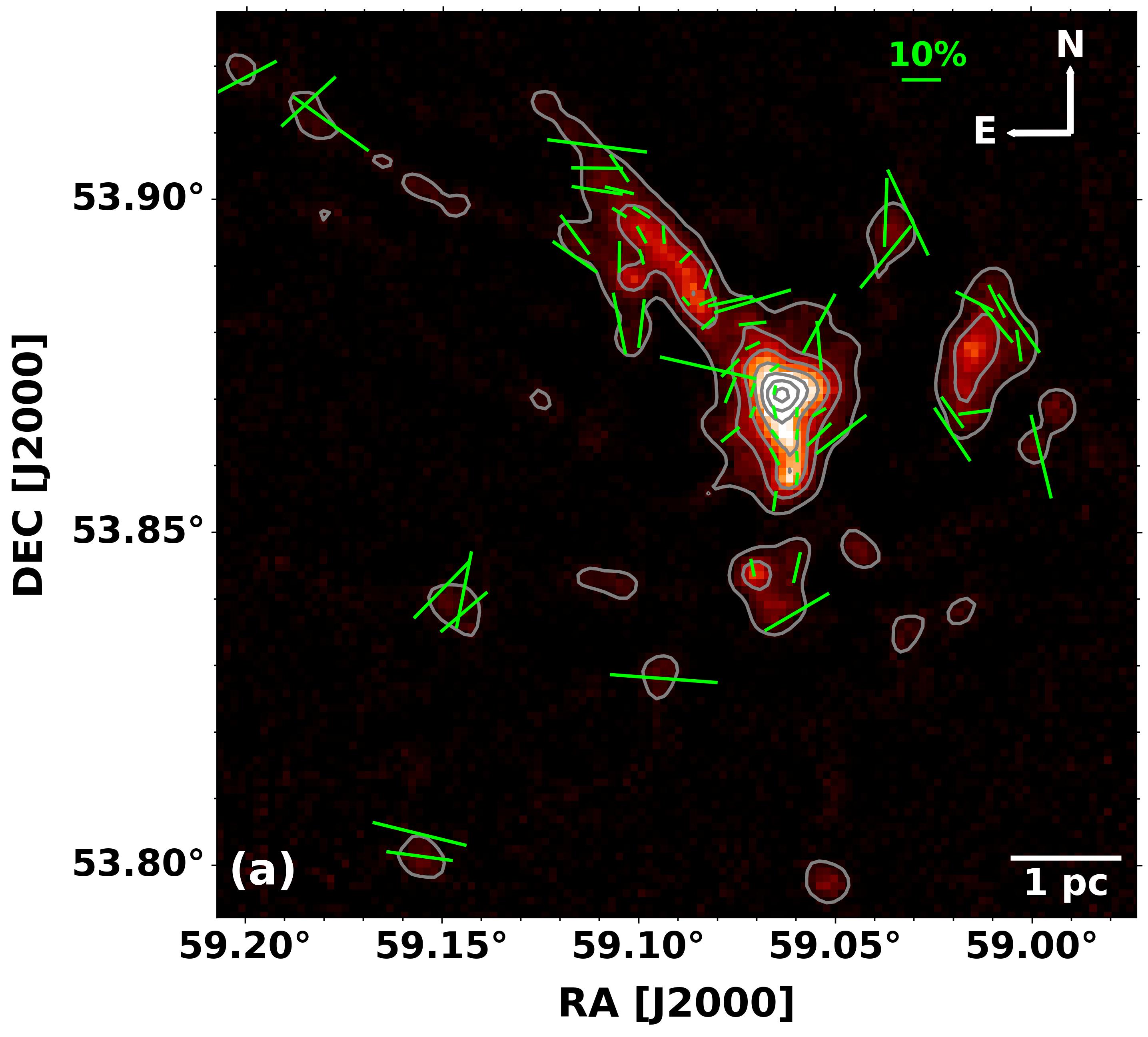}
    \includegraphics[width=8.56cm]{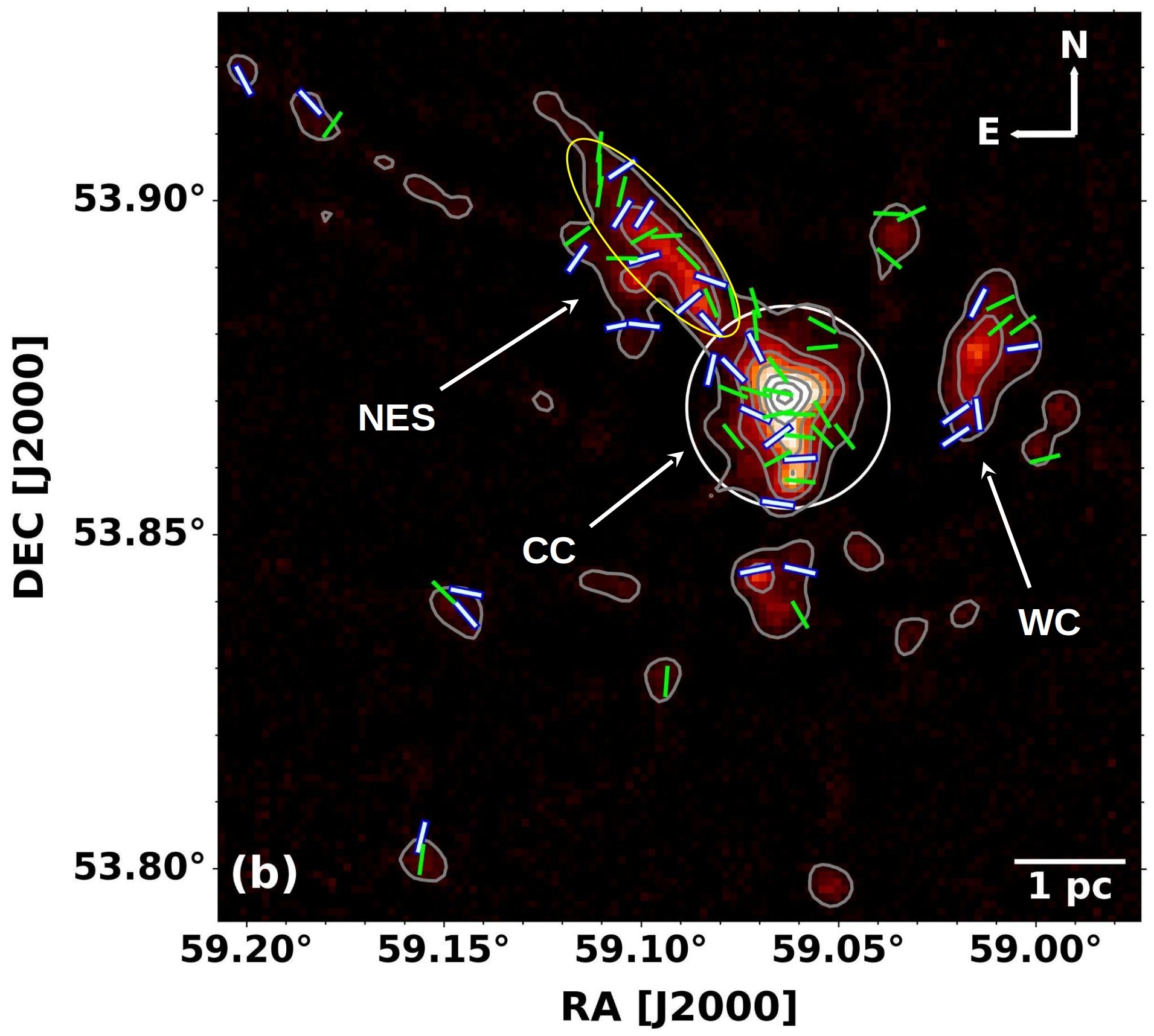}
    \caption{(a) Polarization vector map with lengths proportional to polarization fraction, and (b) magnetic field orientation map with fixed lengths. The background is the Stokes I image at 850 \mum, and the contour levels are drawn at 3$\sigma$ above the rms noise level of 5 \mjb, starting from 15 \mjb to 300 \mjb. The segments shown are binned to a 12\arcsec pixel grid and correspond to polarization data with I/$\delta$I $>$ 10 and P/$\delta$P $>$ 2. The lightcyan and green vectors in panel-b show the measurements with 2 $<$ P/$\delta$P $<$ 3 and P/$\delta$P $>$ 3, respectively. The regions used for B-field calculation are also shown in panel-b by a white circle and yellow ellipse for the CC and NES, respectively.} 
    \label{fig:polmap}
\end{figure*}

In order to select the significant polarization detections, we set the following criteria for selecting data: I/$\delta$I $>$ 10, P/$\delta$P $>$ 2, and P $<$ 30\%. By doing this, we got 69 polarization measurements in our target region. The $P$ values range from $\sim$2 \% to $\sim$29 \% with a mean and standard deviation around $\sim$11 $\pm$ 8 \%. The B-field orientations are widely distributed, ranging from $\sim$6\degree~to 180\degree~with a mean and standard deviation around $\sim$91\degree $\pm$ 48\degree, suggesting a complex B-field morphology in the region. The mean uncertainties in polarization fraction and polarization angle are $\sim$3.5\% and $\sim$9\degree, respectively. Fig. \ref{fig:polmap}a and b show the distribution of polarization vectors and B-field orientations, respectively, over the 850 \mum Stokes I dust continuum emission map of the region. The contour levels in the map are shown above 3$\sigma$ from the background, where $\sigma$ is the mean rms noise (5 \mjb) of the Stokes I map. 

In this work, based on 850 \mum Stokes I intensity and magnetic field orientations, we defined the central clump, clump located on the western side, and northeastern elongated structure as CC, WC, and NES, respectively, as marked in Fig. \ref{fig:polmap}b. The approximate extents of these regions are defined by considering the outermost closed contours of the 850 \mum Stokes I map.  
From Fig. \ref{fig:polmap}b, it can be seen that the B-field orientations in the CC are mostly oriented along the east-west direction (PA $\sim$90\degree), while some of them are at smaller position angles. There exist mixed B-field orientations in the NES region, some in the low-density area are nearly perpendicular to the major axis of the NES, while some closer to the CC are parallel to it.  
In the WC, most of the B-fields are converging towards the centre, aligned along the southeast direction, which may be influenced by gravity (see Section \ref{local_gravity} and \ref{U_shape}). 
Overall, the B-field morphology around the central region of \cloud~is complex, which is probably due to hierarchical fragmentation and a network of filamentary flows towards the hub, as found by \cite{Rawat_2023} and \cite{Rawat_2024}. 

\begin{figure*}
    \centering
    \includegraphics[width=7.3cm]{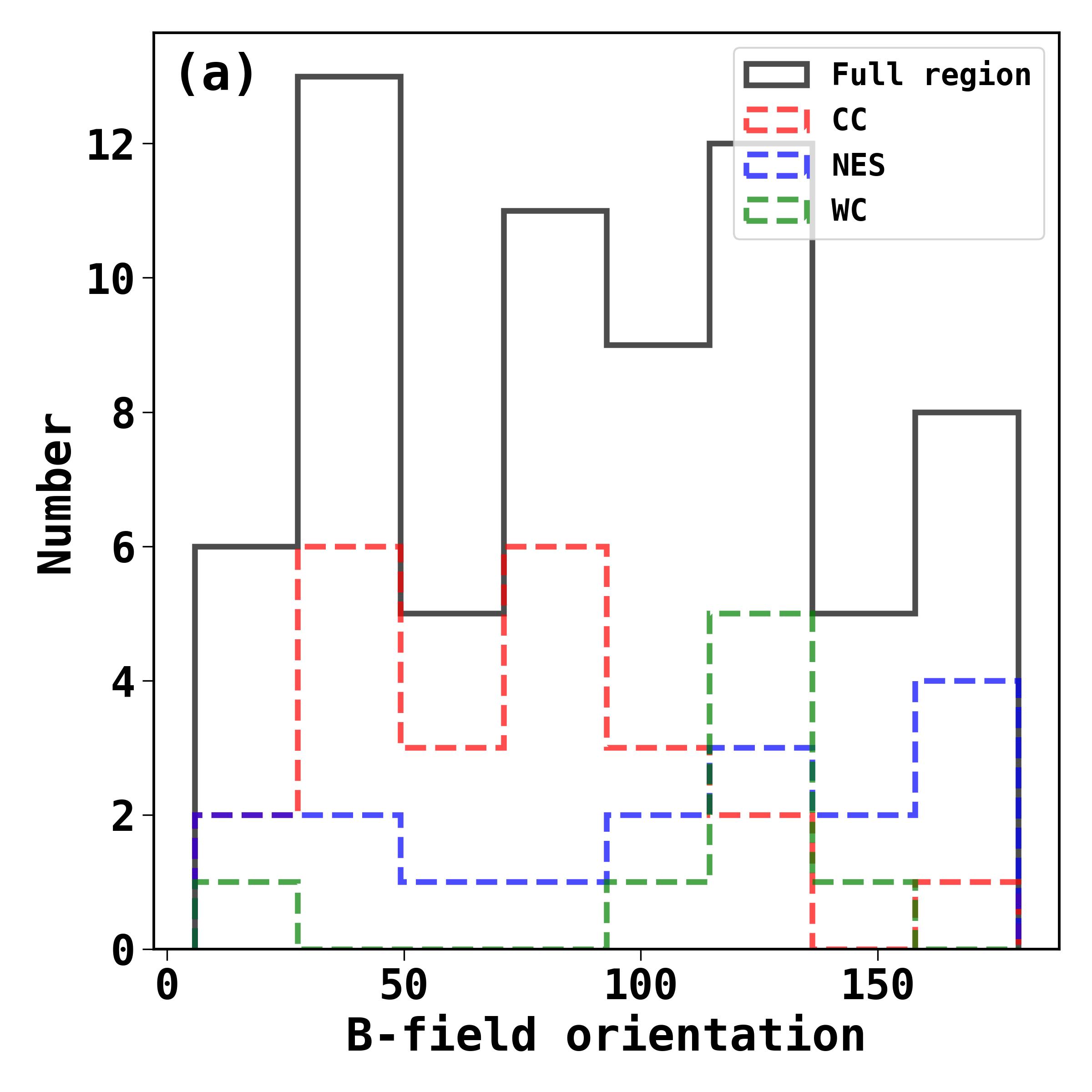}
    \includegraphics[width=9cm]{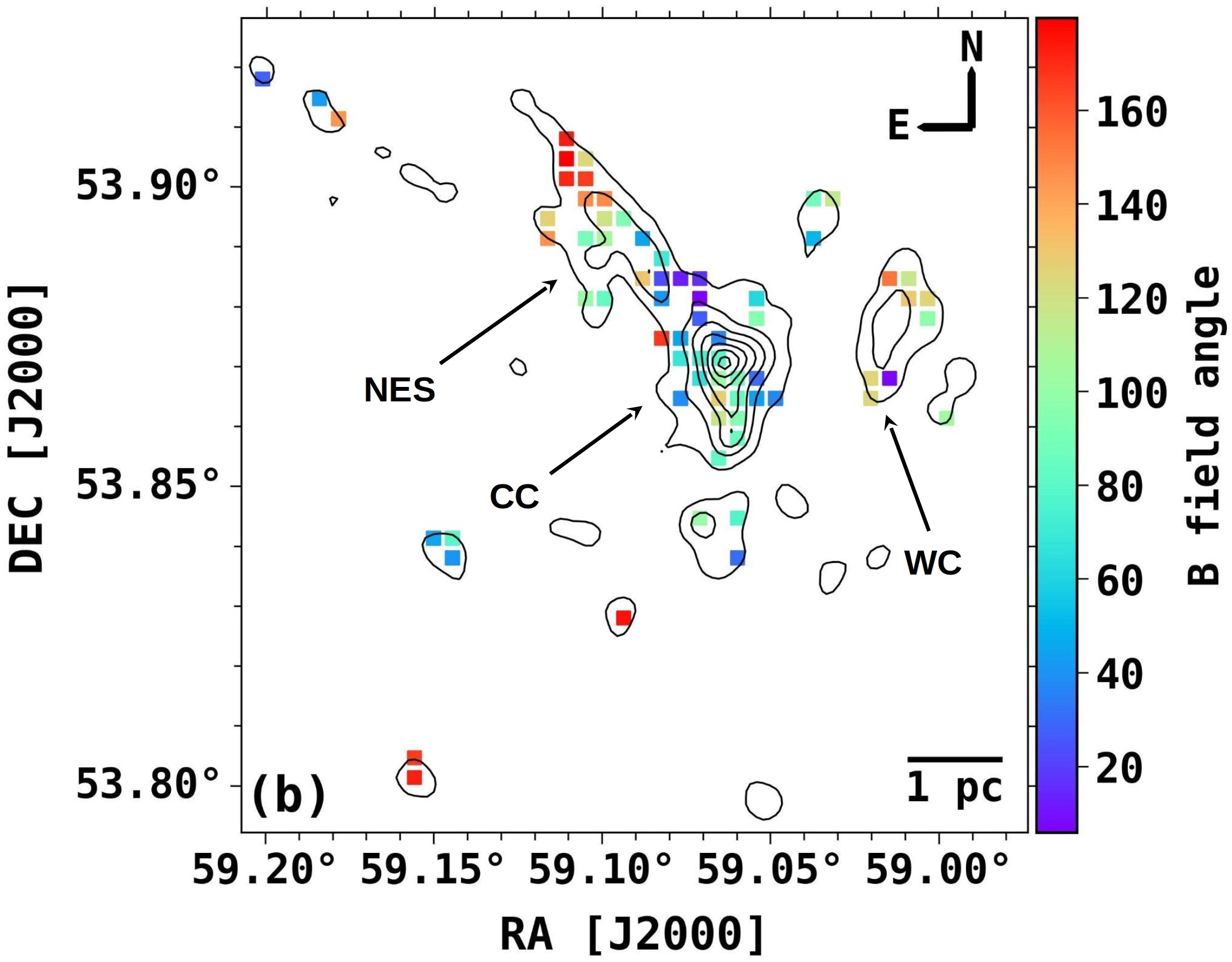}
    \caption{(a) Histogram of B-field position angles for the whole region, CC, NES, and WC. (b) Distribution of B-field position angles over the contours of 850 \mum Stokes I map. The contour levels are the same as in Fig. \ref{fig:polmap}.}
    \label{fig:hist}
\end{figure*}

Fig. \ref{fig:hist}a shows the histogram of the B-field orientations in the central region of \cloud, which is broadly distributed. The CC is showing mixed morphology, having two peaks, one at $\sim$38\degree(i.e. with position angles close to northeast), and the second is at $\sim$80\degree~(i.e. with position angles parallel to east). The NES shows a flat distribution over a broad range, but a slightly higher distribution at a position angle around $\sim $180\degree. The WC, though, has a small number of segments, shows a peak around $\sim $125\degree, i.e. mostly in the southeast direction. 
All these orientations are also clearly evident in Fig. \ref{fig:hist}b, which shows the distribution of B-field position angles.  
From Fig. \ref{fig:hist}b, it can be seen that the B-field angles change roughly from $\sim$180\degree~to $\sim$70\degree~while going from the elongated structure towards the central clump.

\subsection{Variation of polarization fraction: depolarization effect}

\begin{figure}
    \centering
    \includegraphics[width=8.5cm]{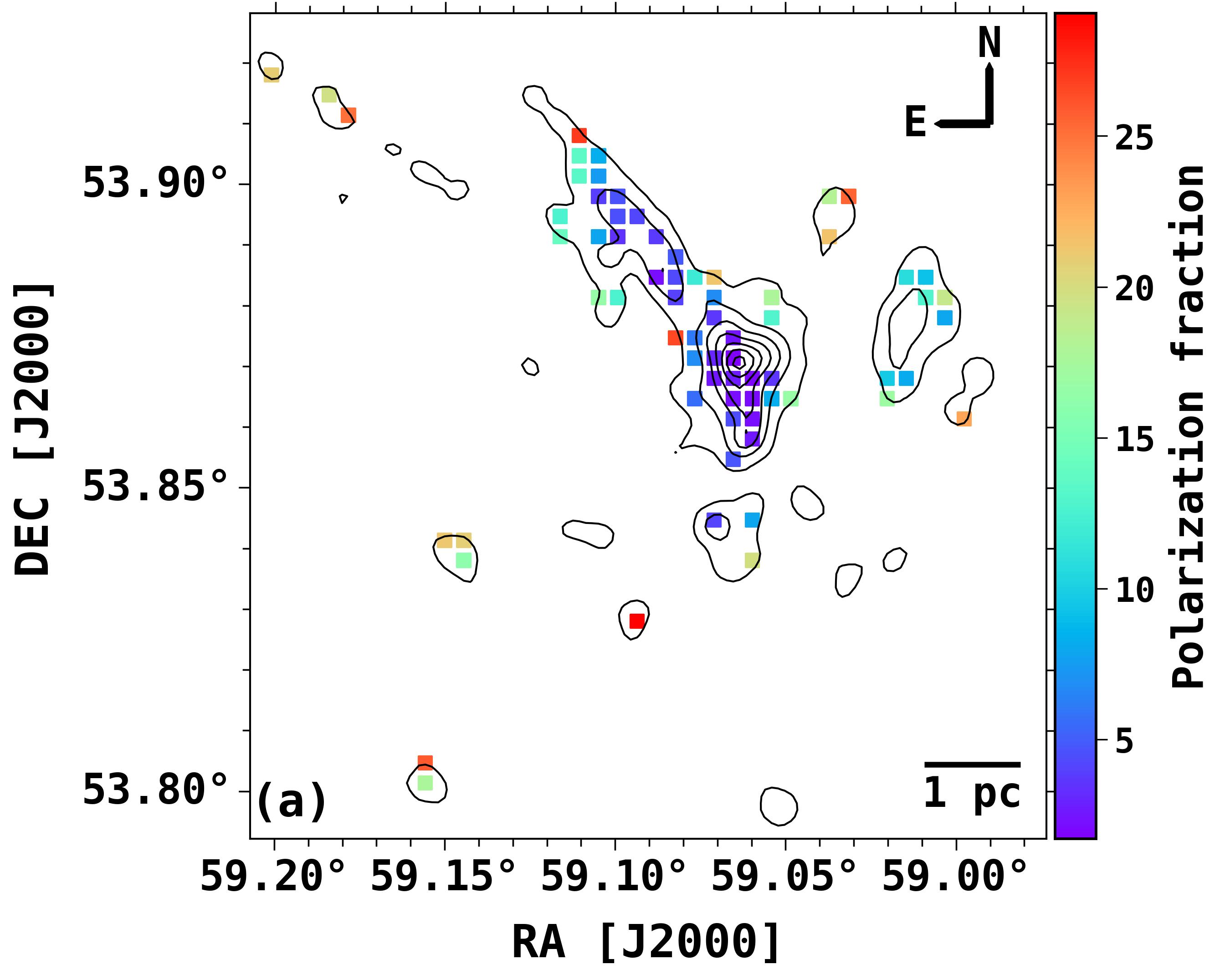}
    \caption{Distribution of dust polarization fraction (P in \%) over the contours of 850 \mum Stokes I map. The contour levels are the same as in Fig. \ref{fig:polmap}.} 
    \label{fig:P_I}
\end{figure}

Fig. \ref{fig:P_I} shows the distribution of polarization fraction over the contours of Stokes I emission. From the figure, it can be seen that the polarization fraction is lower in the high-intensity regions compared to the low-intensity regions, 
which shows the decreasing trend of polarization fraction with the total intensity, known as depolarization. The depolarization effect has been reported in several studies \citep{Girart_2006, Tang_2013, Sadavoy_2018, Soam_2018, Liu_2019, Liu_2020}, and is mainly explained by the inefficient radiative alignment of dust grains in high-density regions or integration effect across a complex magnetic fields. In high-density regions, the radiative alignment torques decrease due to the attenuation of interstellar radiation that results in poor grain alignment, and hence the decrease in polarization fraction. However, the grain characteristics like size, shape, composition, and grain growth can also affect the dust grain alignment. The turbulent nature of the B-field and unresolved complex and tangled B-fields within the JCMT beam, being averaged across the beam, can also give low dust polarization \citep{Planck_2016, Planck_2020}. We want to point out that in the hub/C1 clump of \cloud, supersonic non-thermal motions have been found (sonic Mach number = 3$-$4.4, see \cite{Rawat_2024}). Therefore, the turbulent nature of the B-field can also be the cause of depolarization in our target.\\   

The relation between polarization fraction and intensity is expected to follow a power-law, $P \propto I^{-\alpha}$ \citep{Whittet_2008}. A range of $\alpha$ values has been found in molecular clouds from $\sim$0.5 to 1 \citep[][and references therein]{Chung_2023}. 
The $\alpha$ value is often used as an indicator of the dust grain alignment efficiency. When $\alpha$ = 0, it implies a constant grain alignment efficiency, $\alpha$ = 0.5 implies that the alignment decreases linearly with the increasing optical depth, while $\alpha$ = 1 implies an alignment limited to the outer regions of the cloud, and at higher density, there is no preferred alignment of grains relative to the magnetic field \citep{Whittet_2008}. We fit the P-I relation with a single power-law (weighted-fit) and found an index, $\alpha$ = 0.95 $\pm$ 0.04, which shows that the dust grain alignment efficiency is decreasing in the central dense region of \cloud. However, \cite{Pattle_2019} shows that the conventional approach of fitting a single power-law over the polarization measurements debiased with Gaussian noise is only applicable above a high SNR cut. But in low polarized intensity regions, a high SNR would discard more data, and therefore, the $\alpha$ index will be overestimated \citep{Pattle_2019, Chung_2023, Lin_2023}.\cite{Wang_2019} also found that the value of $\alpha$ depends upon the cut of SNR and tends to $-1$ if we put a constraint on P/$\delta$P. Hence, to obtain the true value of the $\alpha$ index, it is recommended to use the non-debiased polarization measurements, including both the low and high SNR data, and should not put constraints on P/$\delta$P \citep{Pattle_2019, Wang_2019}.

\begin{figure}
    \centering
    \includegraphics[width=8.5cm]{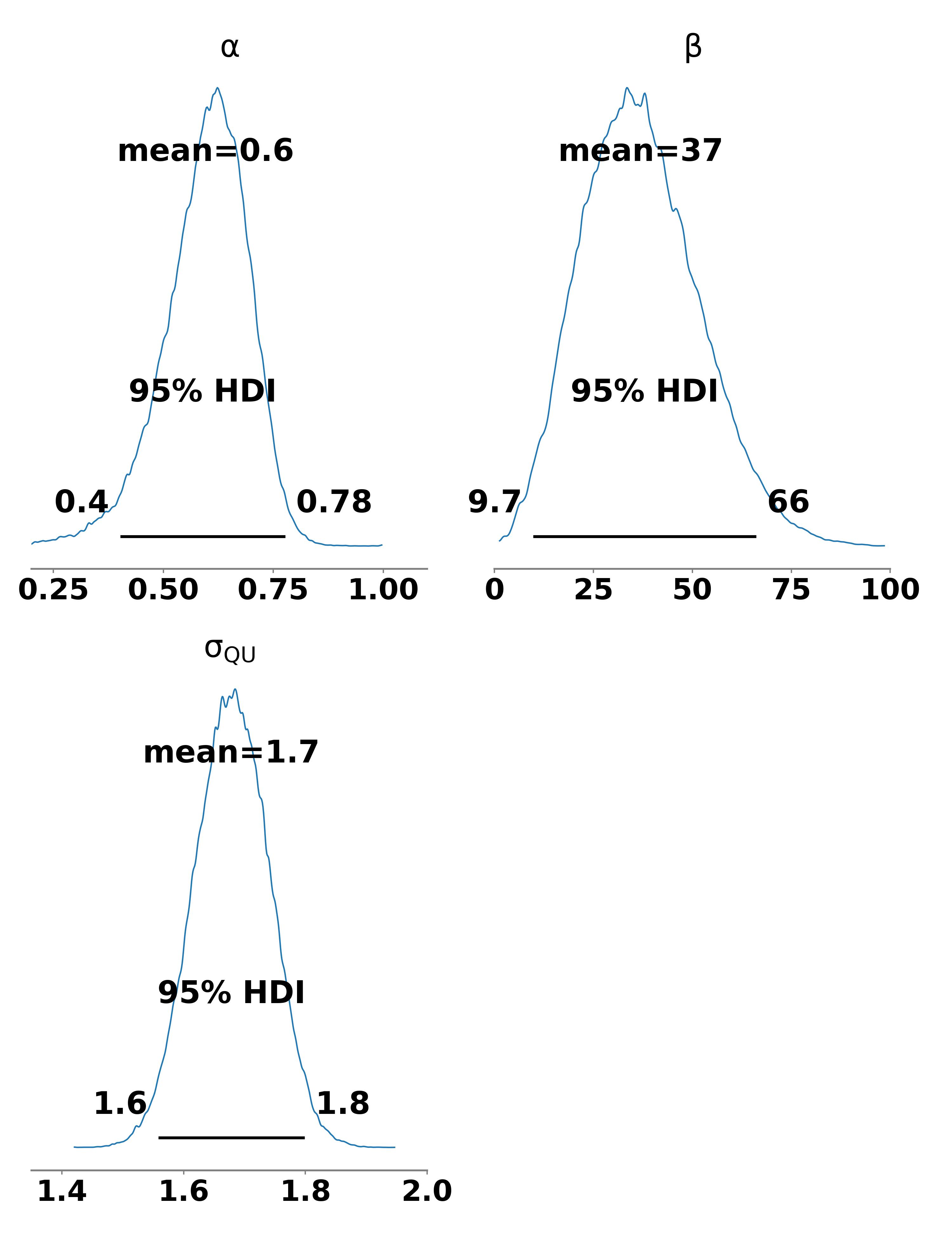}
    \caption{The probability distribution function of the fitted model parameters derived using the Bayesian method over the non-debiased polarization data. The mean values of the parameters are shown along with the 95\% HDI intervals to represent the uncertainties. The 95\% conﬁdence intervals are marked as horizontal bars.}
    \label{post_hist} 
\end{figure}

\begin{figure}
    \centering
    \includegraphics[width=8.5cm]{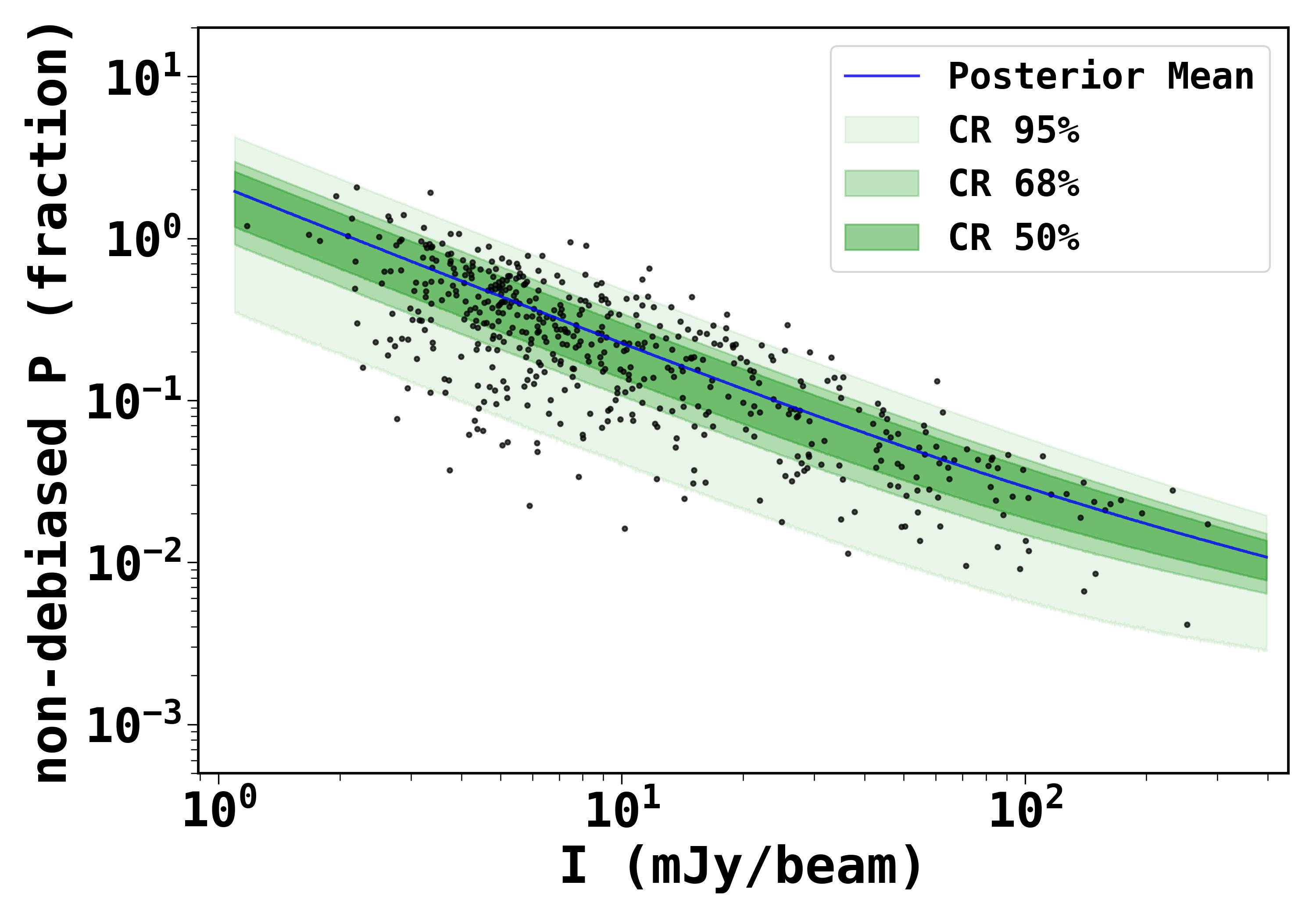}
    \caption{Non-debiased polarization fraction versus total intensity. The blue line shows the mean, and the coloured regions show the 95\%, 68\%, and 50\% confidence limits, as predicted by the posteriors of $\alpha$ = 0.6, $\beta$ = 37, and $\sigma_{QU}$ = 1.7.}
    \label{post_interval} 
\end{figure}

We followed the Bayesian method of \cite{Wang_2019} to determine the true value of $\alpha$ by using the non-debiased polarization data, which follows well the Rice distribution \citep[see][and references therein]{Pattle_2019, Wang_2019} 

\begin{equation}
    F(P|P_0) = \frac{P}{\sigma_P^2} \exp{\left[-\frac{P^2 + P_0^2}{2\sigma_p^2}\right] I_0 \left(\frac{PP_0}{\sigma_P^2}\right)},
\label{bayesian}
\end{equation}
where $P$ and $P_{\rm{0}}$ are the observed and true polarization fraction, respectively, $\sigma_{P}$ is the uncertainty in the polarization fraction, and $I_0$ is the zeroth-order modiﬁed Bessel function. We used non-debiased data with an SNR of 2 (i.e. I/$\delta$I $>$ 2) to include most of the data points and used the power-law model, $P_{\rm{0}} = \beta I^{-\alpha}$, with uncertainty $\sigma_P = \sigma_{\rm{QU}}/I$, where $\sigma_{\rm{QU}}$ represents the rms noise in Q and U measurements, $I$ is the total observed intensity, and $\alpha$, $\beta$, and $\sigma_{\rm{QU}}$ are the free model parameters. We employed the Markov Chain Monte Carlo method and used a python package PyMC3 \citep{Salva_2016} to fit the Rician model to the data. We set the uniform priors on all three model parameters: 0 $<$ $\alpha$ $<$ 2, 0 $<$ $\beta$ $<$100, and 0 $<$ $\sigma_{\rm{QU}}$ $<$ 5, and otherwise a value of 0 for all the parameters. The details of the methodology are given in \cite{Wang_2019}. Fig. \ref{post_hist} shows the derived posterior of each model parameter, along with their 95\% highest density interval (HDI), depicting the uncertainty in each parameter. The mean values of $\alpha$, $\beta$, and $\sigma_{QU}$ are $\sim$0.6, 37, and 1.7, respectively. The $\alpha$ value derived from the non-debiased polarization data is smaller than the $\alpha$ value derived from the conventional approach (i.e. $\sim$0.95). Fig. \ref{post_interval} shows the non-debiased polarization fraction versus total intensity plot with 50\%, 68\%, and 95\% confidence interval. The derived $\alpha$ value suggests that the grain alignment is still persisting in the hub of \cloud, but with decreasing efficiency in the dense regions. 

\subsection{Relative orientations of magnetic fields, intensity gradients, and local gravity}
\label{comparison_B_IG_LG}
In star-forming regions, various forces interact, like gravity, magnetic field, and turbulence, which shape the geometry of these regions and drive the star-formation process \citep{Balles_2007, Koch_2012a, Pattle_2022}. 
Along with the overall strength of these individual factors for an entire region (discussed in Section \ref{comp}), it is also important to investigate their localised relative orientations in the map, as it would give insight into the localised effect of these factors \citep{Koch_2012a, Koch_2012b, Koch_2013, Tang_2019, Liu_2020, Wang_2020b}. \cite{Koch_2012a} developed a technique, "the polarization-intensity gradient-local gravity," using Magnetohydrodynamics (MHD) force equations to measure the local magnetic field strengths. Following the approach of \cite{Koch_2012a, Koch_2012b}, we find out the angular difference between magnetic field, intensity gradient, and local gravity, and discuss their relative importance at different positions.

\subsubsection{Intensity gradient versus magnetic field}
\label{B_vs_IG}
We used the 850 \mum dust continuum intensities of all pixels in the map to determine the directions of intensity gradients. All the pixels that have values above a certain threshold (i.e. 3$\sigma$ above the mean rms noise in the Stokes I map) are considered for computing the direction of gradients, except those that exist at the edges. For a pixel at position ($\alpha_i$, $\delta_j$), the position angle (${\theta ^\prime}_{IG}$) of the intensity gradient is calculated as

\begin{equation}
    {\theta ^\prime}_{IG} = (180/\pi) \times \arctan{\left[\frac{\Delta I_{\delta_j}}{\Delta I_{\alpha_i}}\right]},
\end{equation} where $\Delta I_{\delta_j} = I_{\delta_{j+1}} - I_{\delta_{j-1}}$ and $\Delta I_{\alpha_i} = I_{\alpha_{i+1}} - I_{\alpha_{i-1}}$.\\

\begin{figure*}
    \centering
    \includegraphics[width=8.0cm]{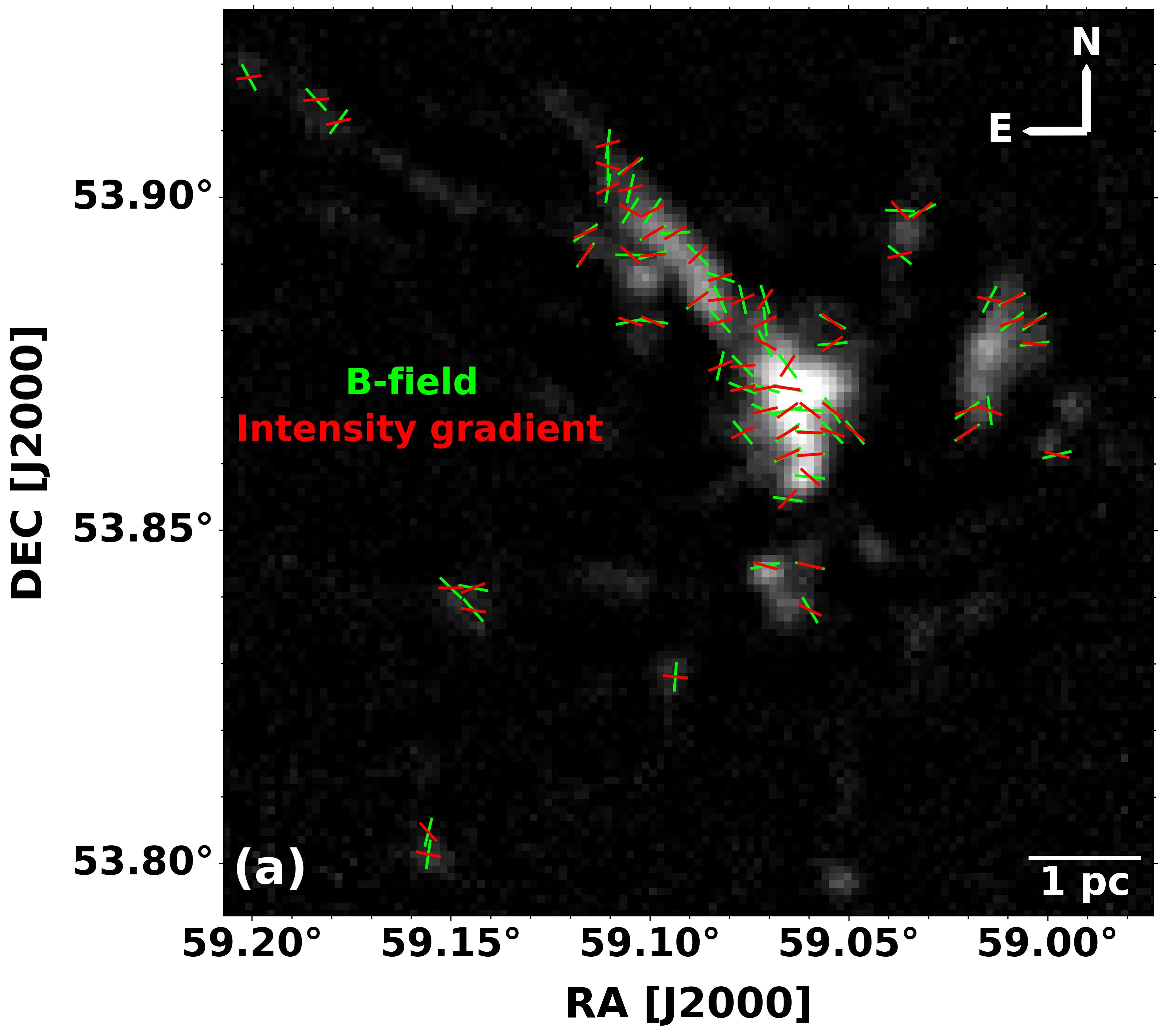}
    \includegraphics[width=8.5cm]{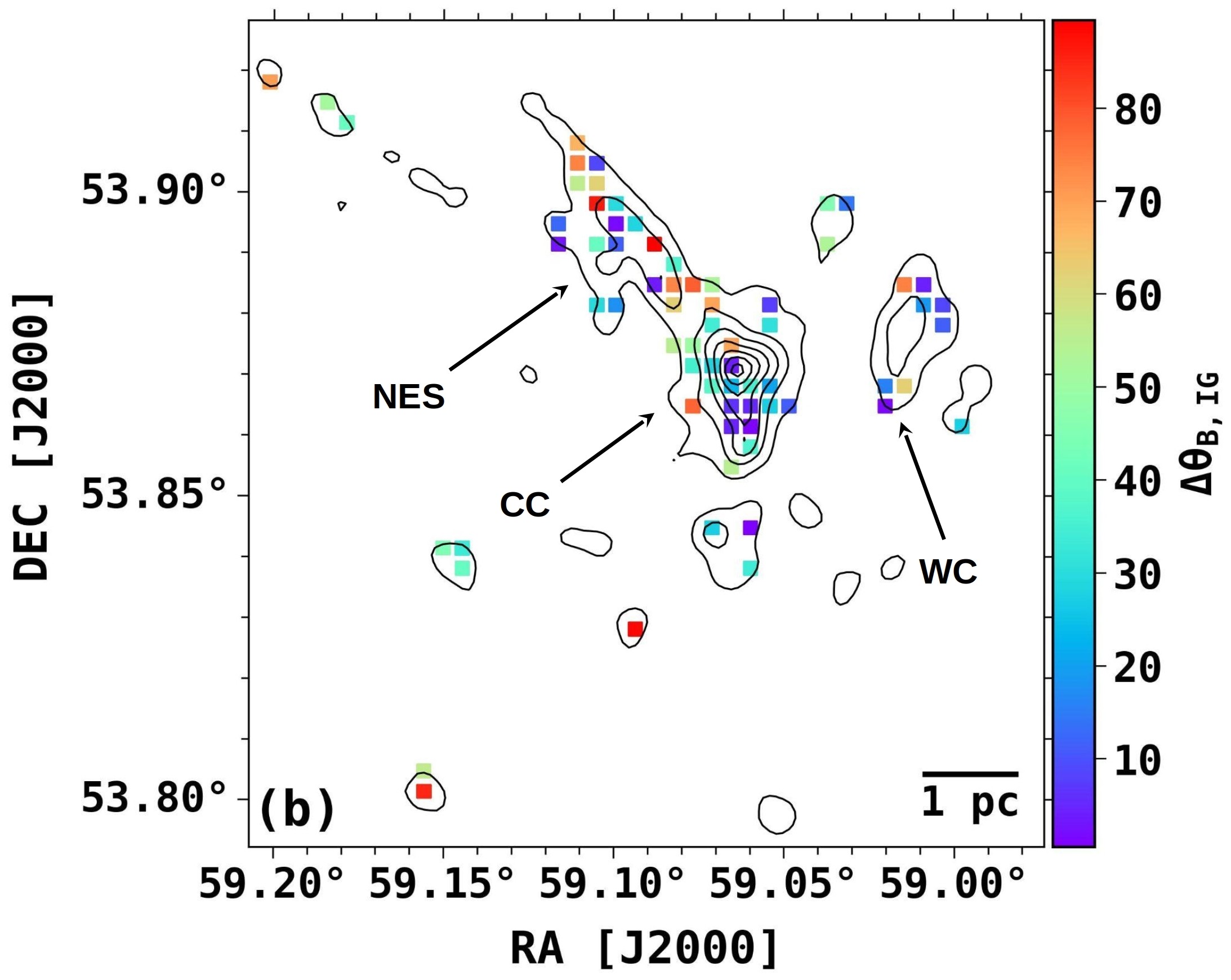}
    \caption{(a) The orientations of the B-ﬁelds (green segments) and intensity gradients (red segments) are overlaid on the 850 \mum Stokes I map. (b) The distribution of the offset between the position angles of the B-ﬁelds and intensity gradients, i.e., $\Delta \theta_{B, IG}$ = $|(\theta_B - \theta_{IG})|$ over the contours of 850 \mum Stokes I map. The contour levels are same as in Fig. \ref{fig:polmap}. }
    \label{fig:B_vs_int}
\end{figure*}

The ${\theta ^\prime}_{IG}$ values are then converted to gradient directions ($\theta_{IG}$) by doing the quadrant corrections, i.e. arranging the angles between 0\degree~and 360\degree~\citep[for details, see][]{Eswaraiahetal2020}. In order to plot the gradient orientations instead of directions, we folded the $\theta_{IG}$ between 0\degree~and 180\degree. For comparison of the gradient orientations ($\theta_{IG}$) with the B-field orientations ($\theta_{B}$), we took the average of all the  $\theta_{IG}$ values within a diameter of $\sim$14\arcsec (corresponds to the beam size of JCMT at 850 \mum) around each B-field position. 
We calculated the circular mean to get the average of intensity gradients. In this approach, the angles are treated as unit vectors, which is adequate for broad distributions and ambiguity in angles \citep{Tang_2019}. Fig. \ref{fig:B_vs_int}a shows the orientations of intensity gradients relative to B-field orientations over the 850 \mum Stokes I map.
\\
We ﬁnd that the local differences between these
orientations are overall widely distributed. However, it can be seen that the intensity gradients are mostly aligned with the B-fields in the CC and WC regions, while the differences in orientations are relatively higher in the NES region. 
In the observed central region of \cloud, we found a moderate correlation between $\theta_B$ and $\theta_{IG}$, in the CC and WC (see Figs. \ref{fig:B_vs_int}a, b). Fig. \ref{fig:B_vs_int}b shows the distribution of $\Delta \theta_{B, IG}$ = $|(\theta_B - \theta_{IG})|$ over the contours of 850\mum Stokes I map. The $\Delta \theta_{B, IG}$ values lie between 0\degree~and 90\degree~after considering them to be the acute angle.
A stronger correlation between $\theta_B$ and $\theta_{IG}$ tells that the material is following the B-field lines 
\citep[][more discussion in Section \ref{U_shape}]{Koch_2013, Tang_2019}. 

\subsubsection{Local gravitational field versus magnetic field }
\label{local_gravity}
\begin{figure*}
    \centering
    \includegraphics[width=8.0cm]{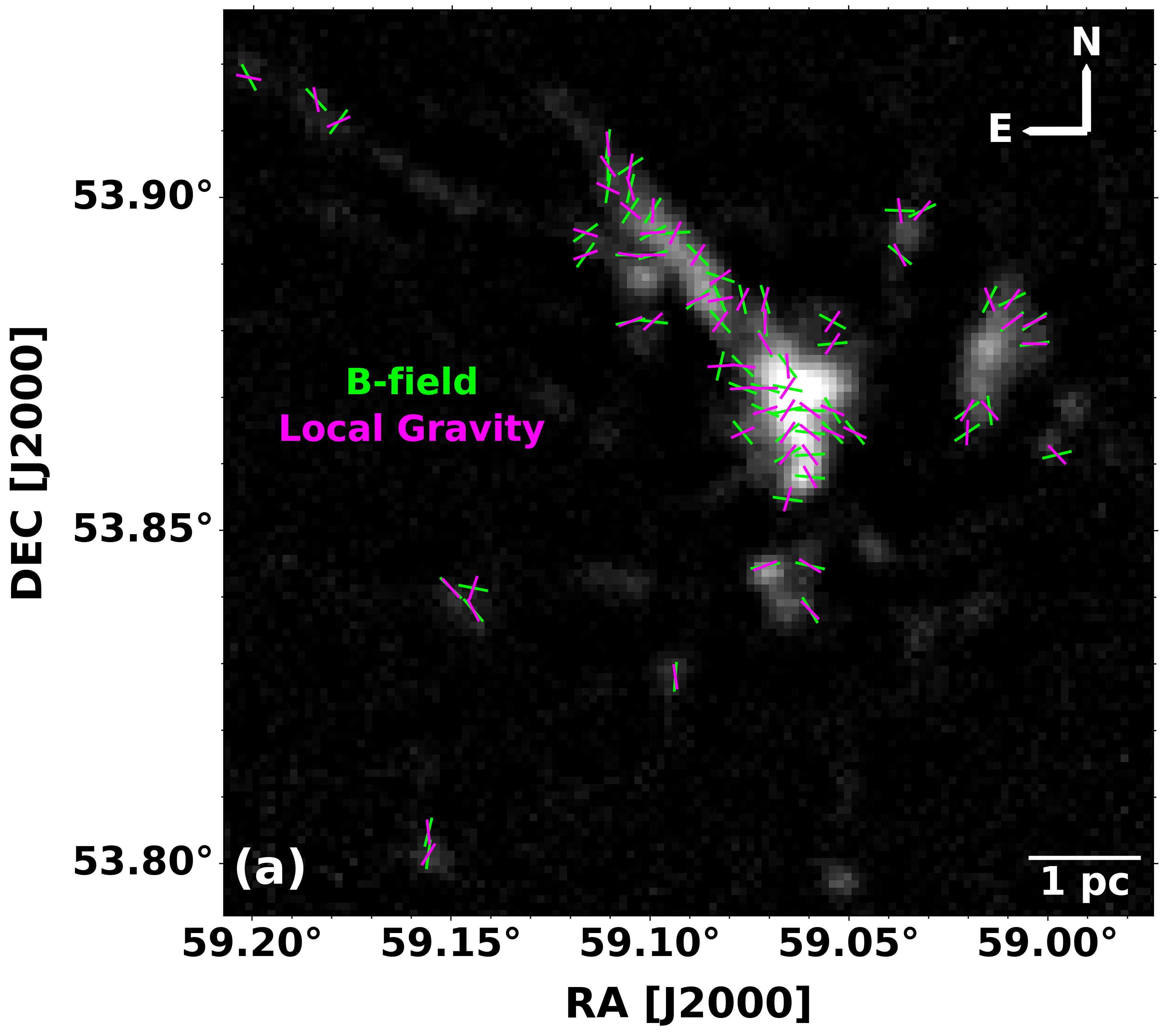}
    \includegraphics[width=8.5cm]{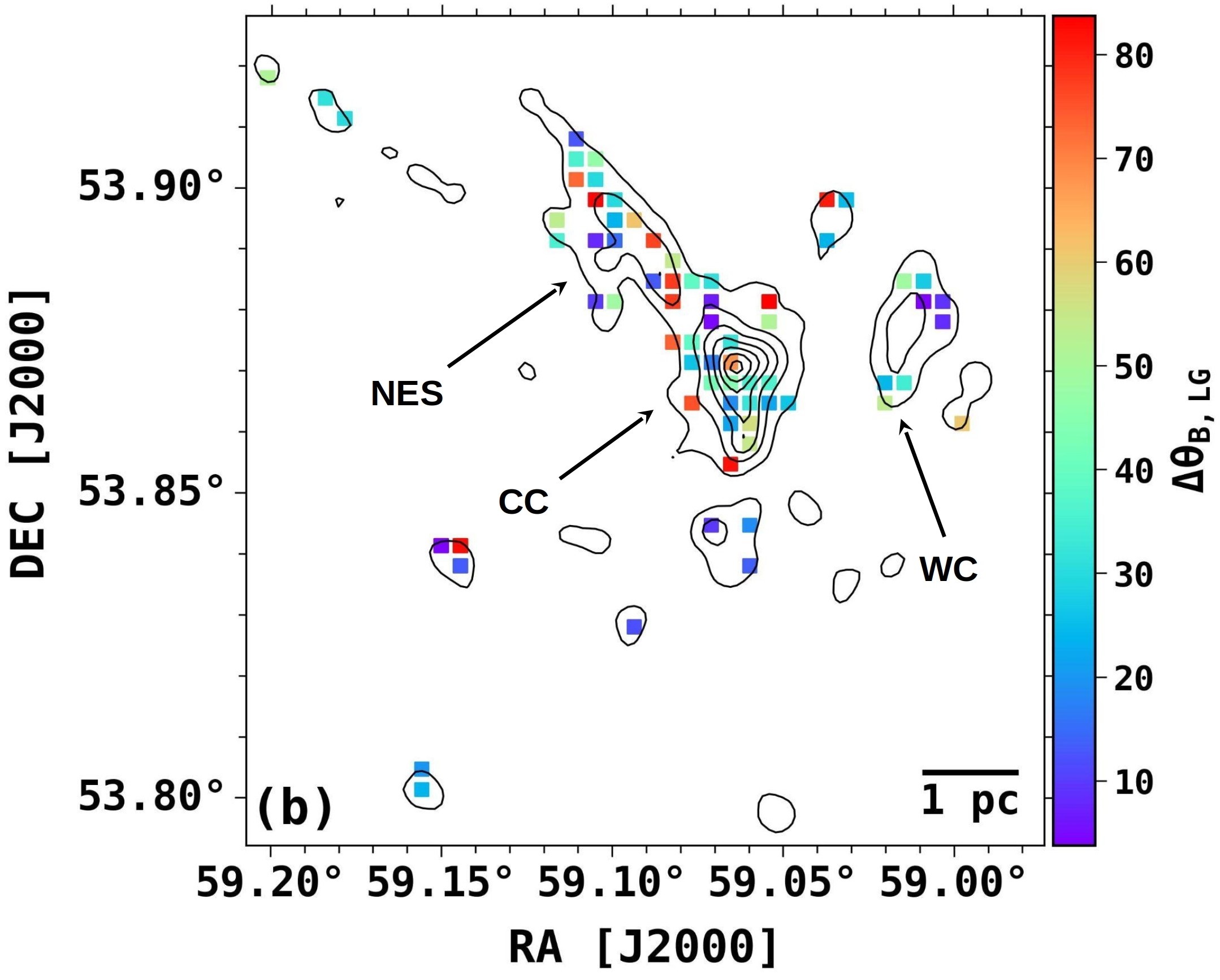}
    \caption{(a) The orientations of the B-ﬁelds (green segments) and local gravity (magenta vectors) are overlaid on the 850 \mum Stokes I map. (b) The distribution of the offset between the position angles of the B-ﬁelds and local gravity, i.e., $\Delta \theta_{B, LG}$ = $|(\theta_B - \theta_{LG})|$ over the 850 \mum Stokes I map. The contour levels are same as in Fig. \ref{fig:polmap}. }
    \label{fig:gravity_map}
\end{figure*}

 In order to investigate the localised effect of gravity on the B-field morphology of the structures, we used the 850 \mum dust continuum map, to compute the projected gravitational field vectors. 
 The gravitational force at any pixel ($F_{G, i}$) is the vector sum of the forces from all the surrounding pixels and is expressed as \citep{Wang_2020b}.

\begin{equation}
    F_{G,i} = kI_i \sum_{j=1}^{N} \frac{I_j}{{r^2}_{ij}} \hat{r},
\end{equation}

\noindent where $I_i$ and $I_j$ are the intensity of the pixel at position $i$ and $j$, respectively, and $k$ is the term that takes care of the conversion of emission to total column density and also includes the gravitational constant. $N$ is the total number of pixels within the selected area, ${r}_{ij}$ is the projected distance between the pixels i and j, and $\hat{r}$ is the unit vector. Considering only the directions of the local gravitational forces, we take $k$ to be 1 in the above equation by assuming that the spatial distribution of dust will be analogous to the spatial distribution of mass. 
Similar to the intensity gradient map, we selected those pixels which have intensity values above the threshold, 
and obtained the local gravity vectors ($\theta_{LG}$) at each B-field position, by taking an average of all vectors within the 14\arcsec beam size. Fig. \ref{fig:gravity_map}a shows the orientations of local gravity vectors relative to B-field orientations over the 850 \mum Stokes I map. From the figure, it can be seen that similar to intensity gradients, the local gravity vectors are also mostly
aligned with the B-fields in the CC and WC region, whereas they deviate from the
B-fields in the NES region. Fig. \ref{fig:gravity_map}b shows the distribution of \textbf{$\Delta \theta_{B, LG}$} = $|(\theta_B - \theta_{LG})|$ values over the contours of 850 \mum Stokes I emission, which are treated to be acute angles.

\begin{figure*}
    \centering
    \includegraphics[width=5.65cm, height = 5.65cm]{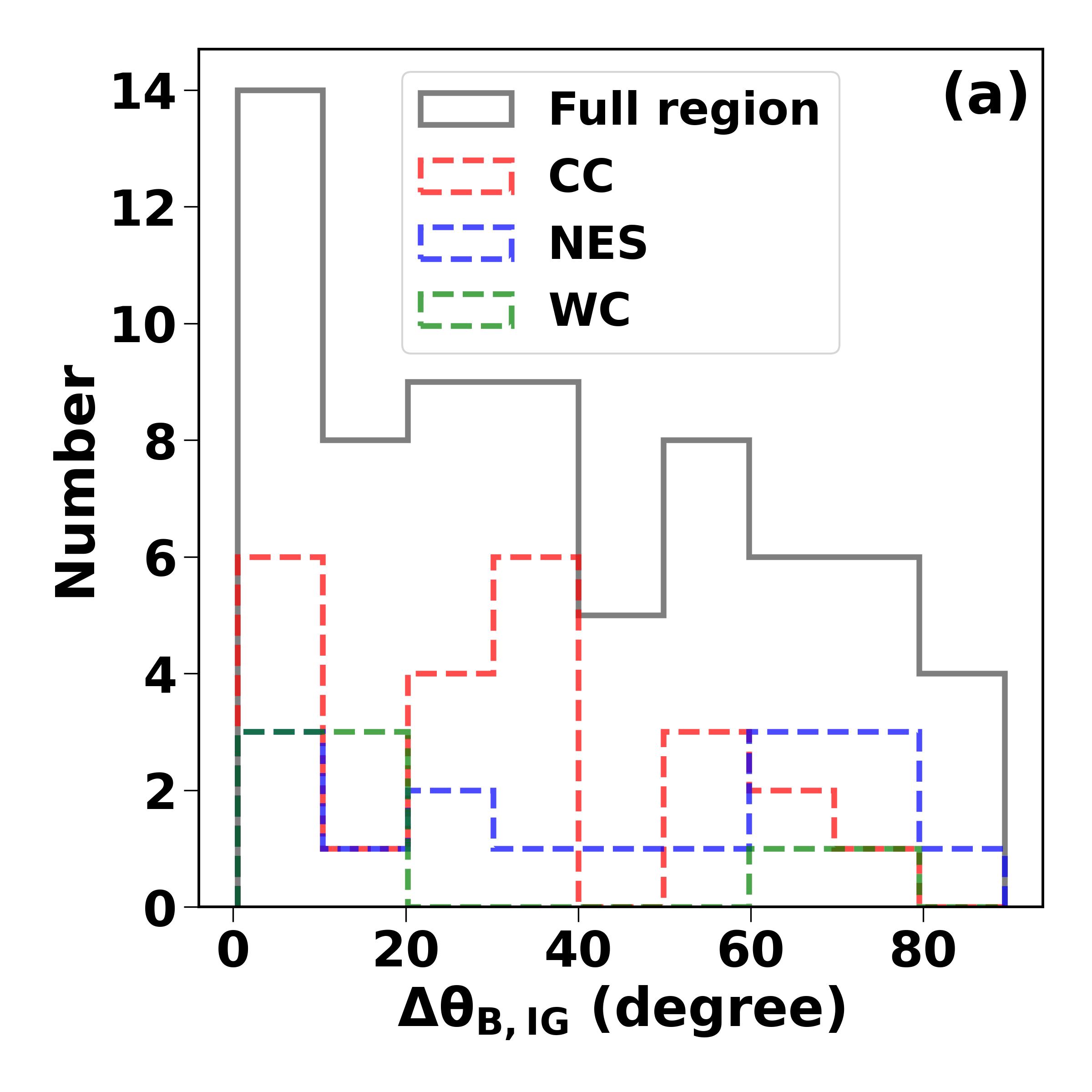}
    \includegraphics[width=5.5cm]{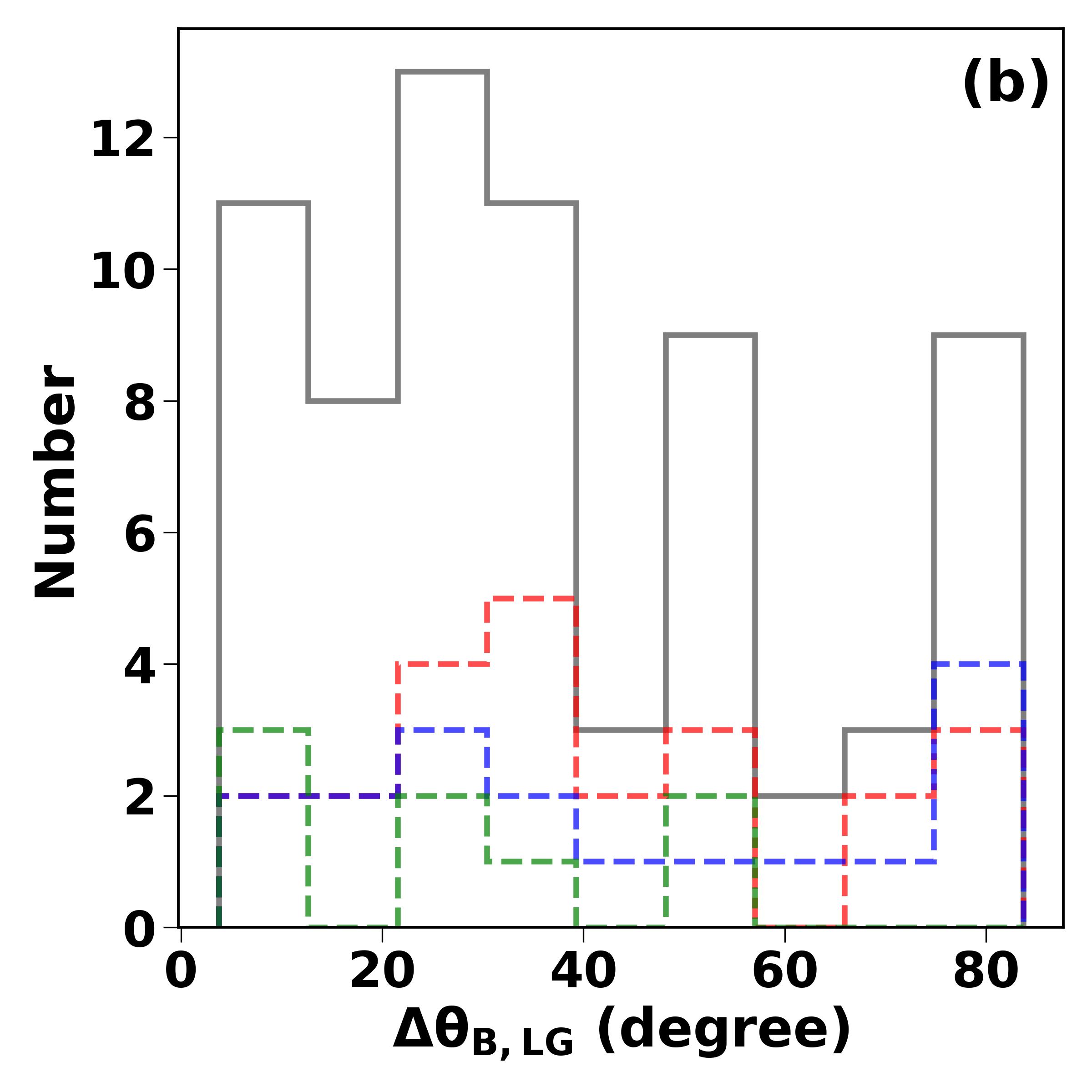}
    \includegraphics[width=5.5cm]{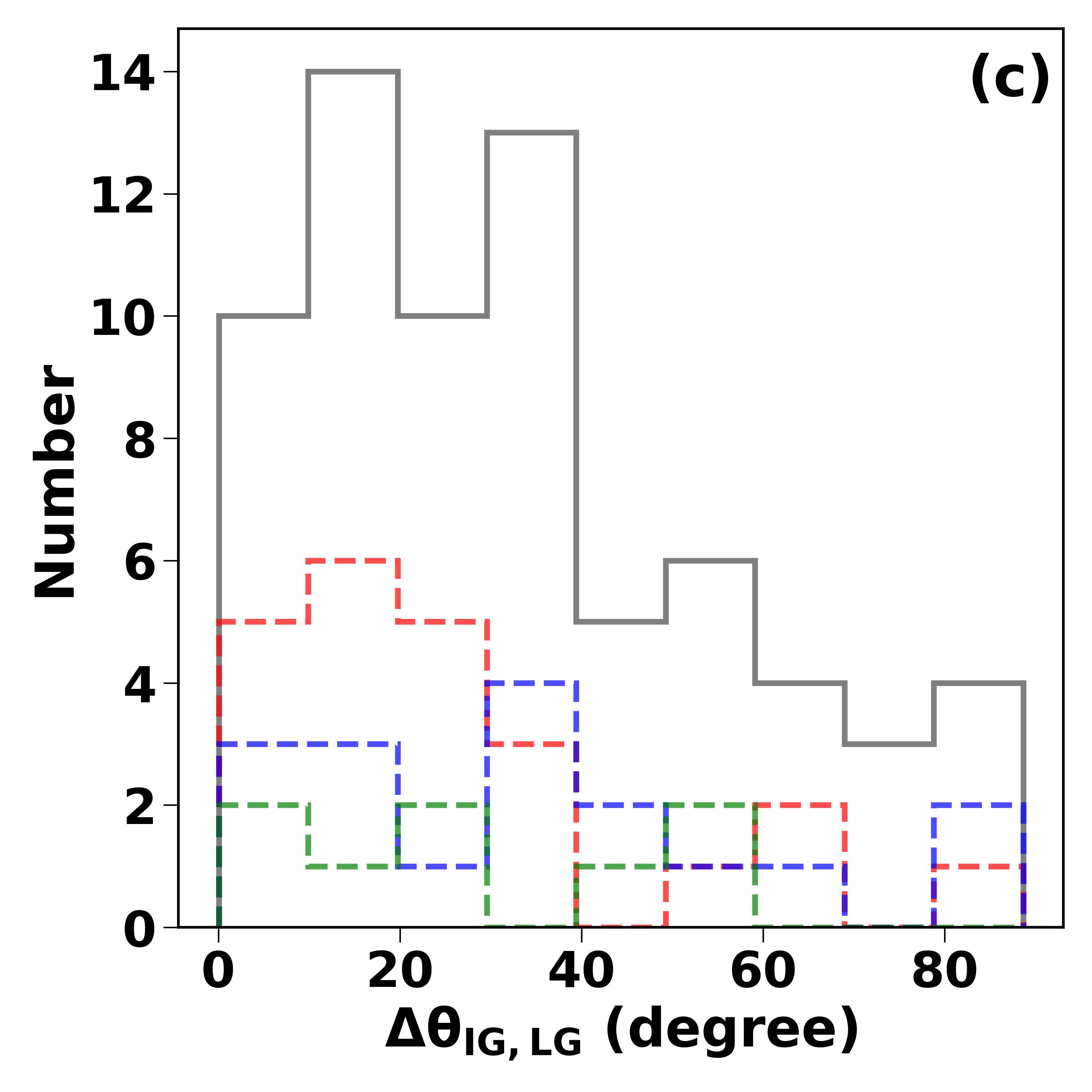}
    \caption{Distribution of difference in position angles of (a) magnetic field ($\theta_B$) and intensity gradient ($\theta_{IG}$), (b) magnetic field ($\theta_B$) and local gravity ($\theta_{LG}$), and (c) intensity gradient ($\theta_{IG}$) and local gravity ($\theta_{LG}$). The red, blue, and green histograms show the difference in position angles within the CC, NES, and WC regions, respectively.}
    \label{fig:hist_comp}
\end{figure*}

The relative orientations of intensity gradient and local gravity have a similar distribution like that of $\Delta \theta_{B, IG}$ and $\Delta \theta_{B, LG}$, as shown in Fig. \ref{fig:IG_vs_LG} in appendix A. Figs. \ref{fig:hist_comp}a-c shows the histogram distribution of the relative position angle differences of B-field, intensity gradients, and local gravity, i.e. $\Delta \theta_{B, IG}$ = $|(\theta_B - \theta_{IG})|$, $\Delta \theta_{B, LG}$ = $|(\theta_B - \theta_{LG})|$, and $\Delta \theta_{IG, LG}$ = $|(\theta_{IG} - \theta_{LG})|$. From the figure, it can be seen that the differences in the offset angles are mostly distributed towards the smaller angles. The median of $\Delta \theta_{B, IG}$, $\Delta \theta_{B, LG}$, and $\Delta \theta_{IG, LG}$ is 34\degree, 32\degree, and 30\degree~with median absolute deviation of 22\degree, 18\degree, and 15\degree, respectively. The higher angular deviations in the histograms are primarily due to position angles in the elongated structures on the northeastern side, as well as from some structures located south of the CC.\\

\begin{figure}
    \centering
    \includegraphics[width=8.5cm]{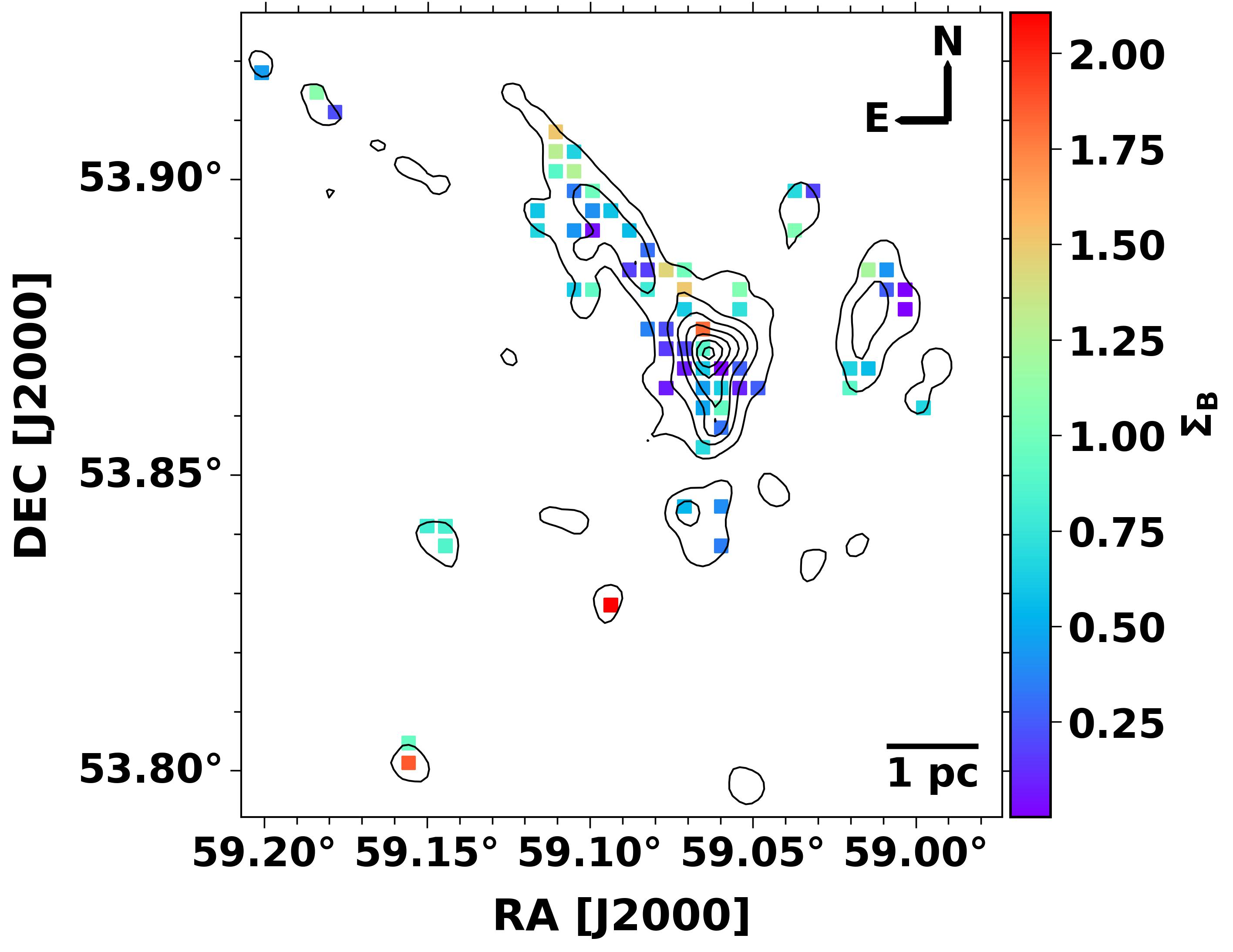}
    \caption{$\Sigma_B$ distribution over the contours of 850 \mum Stokes I map. The contour levels are same as in Fig. \ref{fig:polmap}.}
    \label{fig:sigma}
\end{figure}

As discussed previously, \cite{Koch_2012a, Koch_2012b} developed a "polarization-intensity gradient
method" that can estimate the local ﬁeld-to-gravity force
ratio $\Sigma_B$. This method is based on the assumption that the emission intensity gradients reflect the direction of matter flow due to the combined influences of magnetic pressure force and gravitational force. Using the MHD force equations and geometrically solving them by incorporating the angle between the magnetic field and intensity gradient ($\Delta_{B, IG}$), and between intensity gradient and local gravity ($\Delta_{IG, LG}$), the magnetic ﬁeld ($F_B$)-to-gravity force ($F_G$) ratio can be obtained as

\begin{equation}
    \Sigma_B = \frac{sin(\Delta_{B, IG})}{sin(90 - \Delta_{IG, LG})} = \frac{F_B}{|F_G|}.
\end{equation}
In the above equation, the hydrostatic gas pressure is assumed to be negligible. Fig. \ref{fig:sigma} shows the $\Sigma_B$ distribution plot over the Stokes 850 \mum intensity map. From the figure, it can be seen that $\Sigma_B$ is mostly $\leq$ 1, with a median around $\sim$0.6, which shows that the magnetic field is not solely enough to balance the gravitational force \citep{Koch_2012a}. This implies that gravity dominates over the magnetic field to govern the gas motion towards the centre. However, we note that the POL-2 images generally filter out large-scale structures, and so here, the intensity gradient only traces the local structure on a 4\arcsec pixel-scale. Therefore, to check the effect of large-scale structures on the intensity gradient and local gravity, we generated similar maps from $\it{Herschel}$ 250 \mum image of \cloud and found that the maps are comparable with the JCMT maps. The use of 250 \mum map is an optimal choice because compared to $\it{Herschel's}$ longer wavelength (i.e. 350 and 500 \mum) bands, its resolution ($\sim$18\arcsec) is comparable to the resolution of the JCMT 850 \mum (14\arcsec) map 
and also it is a better tracer of cold dust compared to $\it{Herschel's}$ shorter wavelength (i.e. 70 and 160 \mum) bands.

Due to the relatively low number of B-field segments in the WC region, we focused our further analysis, like the study of structure, dust properties, and B-field strength calculation, towards the CC and NES regions.

\subsection{Column and number densities}

We calculate the molecular hydrogen column density using the 850\mum dust continuum emission. Assuming the dust emission to be optically thin, the column density can be calculated using the relation \citep{Kauffmann_2008}, 

\begin{multline}
    N(H_2) = 2.02 \times 10^{20} \cms \left(e^{1.439\left(\frac{\lambda}{\mathrm{mm}}\right)^{-1} \left(\frac{T_d}{10 K}\right)^{-1}} - 1\right) \\ \times \left(\frac{\kappa_\nu}{\mathrm{0.01 cm^{2} g^{-1}}}\right)^{-1} \left(\frac{S_\nu}{\mathrm{mJy \ \ beam^{-1}}}\right)  \left(\frac{\theta_{HPBW}}{10\arcsec} \right)^{-2}   \left(\frac{\lambda}{\mathrm{mm}}\right)^3,
\end{multline} 
where $S_\nu$ is the flux density in Jy at frequency $\nu$, $\lambda$ is the wavelength (0.85 mm), $T_d$ is the mean dust temperature in K, dust opacity $\kappa_\nu$ = 0.1($\nu$/1 THz)$^\beta$ = 0.0125 cm$\rm{^2 g^{-1}}$ for $\nu$ = 0.353 THz and dust opacity index, $\beta$ = 2 \citep{Batt_2011, Dehar_2012}, and $\theta_{\mathrm{HPBW}}$ is the beam size (14\arcsec at 850 \mum).\\


The mean $T_d$ is taken from the $\it{Herschel}$ Hi-GAL survey \citep{Molinari_2010} dust temperature map \citep{Schisano_2020} of the \cloud~region. Within the boundaries of CC and NES, the mean $T_d$ is around $\sim$16.8 K and 13.0 K, respectively. The total column density, $\sum N(H_2)$, for CC and NES, are found to be (3.4 $\pm$ 1.3) $\times$ 10$^{24}$ \cms and (1.9 $\pm$ 0.7) $\times$ 10$^{24}$ cm$^{-2}$, respectively. Then, assuming the spherical geometry for CC, its number density ($n_{H_2}$) can be estimated using the relation

\begin{equation}
    n_{H_2} = \frac{M}{V\mu m_H} = \frac{\sum N(H_2) \times A_{pixel}}{\frac{4}{3} \pi R_{eff}^3},
\label{eq:ndens}
\end{equation} where $M$ and $V$ are the mass and volume of the region, respectively, $\mu$ is the mean molecular weight assumed to be 2.8 \citep{Kauffmann_2008}, $m_H$ is the mass of hydrogen, $A_{pixel}$ is the pixel area, and $R_{eff}$ is the effective radius. The $R_{eff}$ for CC is calculated as $(Area/\pi)^{0.5}$, and is around $\sim$0.9 $\pm$ 0.1 pc. Though NES is assumed as an elliptical structure with a semi-minor axis, $r_1$ = 0.35 $\pm$ 0.03 pc and a semi-major axis, $r_2$ = 1.10 $\pm$ 0.09 pc (see Fig. \ref{fig:polmap}a), in 3-dimension, this elongated structure could be better described by a cylindrical geometry. Therefore, to calculate the number density of NES, we adopted its radius ($r$) and length ($L$) to be $r_1$ and 2$r_2$, respectively. Under this approximation, we estimated the $n_{H_2}$ for NES by using $V = \pi r^2 L$ in equation \ref{eq:ndens}. The gas mass within the regions is estimated from the total integrated molecular hydrogen column density, using equation 1 given in \cite{Rawat_2023}.  The total column density, $n_{H_2}$, and the mass of the regions are given in Table \ref{tab:results}. The uncertainties in the estimated cloud parameters are mainly due to uncertainty in the gas-to-dust ratio (23\%), the dust opacity index (30\%), and the distance of the cloud (9\%) \citep[][and references therein]{Rawat_2023}. 

\subsection{Velocity dispersion}

\begin{figure}
    \centering
    \includegraphics[width=6cm]{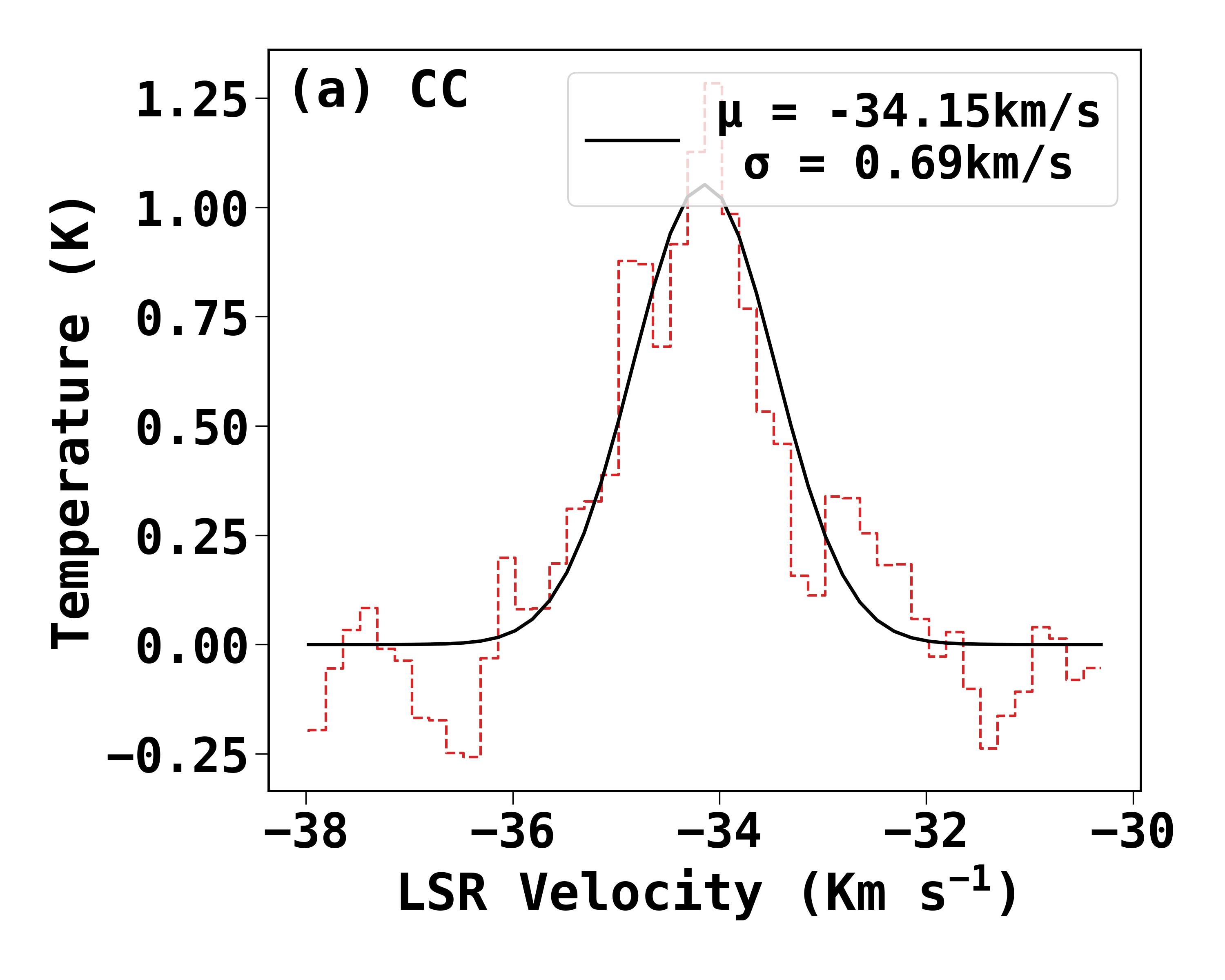}
    \includegraphics[width=6cm]{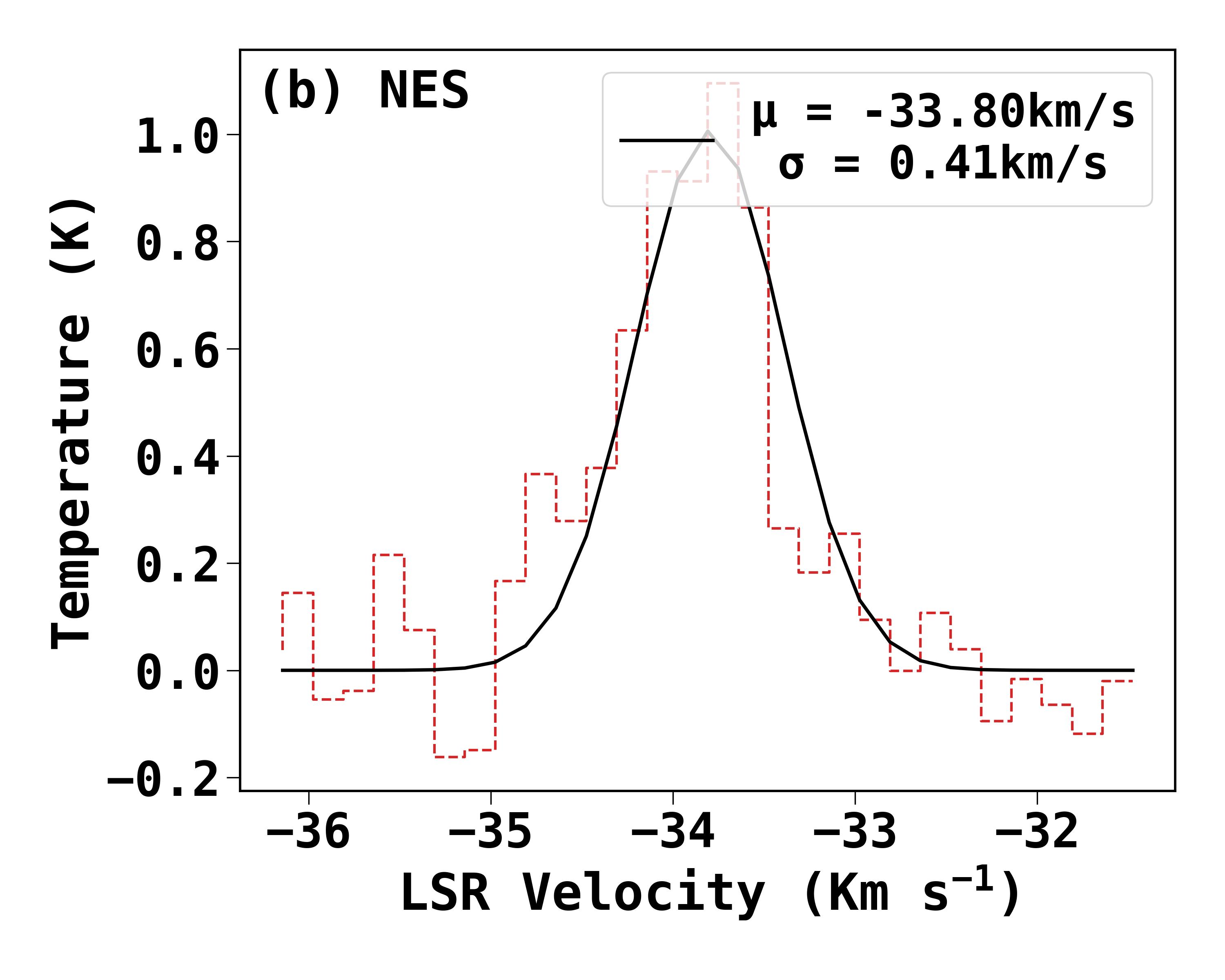}
    \caption{\eico average spectral profile for (a) CC (central clump) and (b) NES (northeastern elongated structure).}
    \label{fig:spectra}
\end{figure}

We used \eico molecular line data to determine the non-thermal velocity dispersion along the line-of-sight. 
\cite{Rawat_2024} found that in comparison to \twco and \thco, \eico emission is optically thin in the \cloud~cloud. Fig. \ref{fig:spectra} shows the \eico spectra averaged within the boundary of the two regions, CC and NES. The Gaussian fitting over the spectra gives the mean velocity as $-$ 34.15 $\pm$ 0.04 \kms and $-$ 33.80 $\pm$ 0.04 \kms, and velocity dispersion ($\sigma_{obs}$) as 0.69 $\pm$ 0.05 \kms and 0.41 $\pm$ 0.03 \kms, for CC and NES, respectively.\\ 

The non-thermal velocity dispersion ($\sigma_{nt}$) can be calculated using the relation, $\sigma_{nt} = \sqrt{\sigma_{obs}^2 - \sigma_{th}^2}$, where $\sigma_{th}$ is the thermal velocity dispersion, defined as $\sqrt{\frac{K_B 
 T_k}{M_{\eico}}}$. Here, $K_B$ is the Boltzmann constant, $T_k$ is the gas kinetic temperature, and $M_{\eico}$ is the mass of \eico molecule (30 amu). We have used the excitation temperature map from \cite{Rawat_2024} to get the approximate values of $T_k$, as 11.4 K and 10.3 K for CC and NES, respectively. The estimated $\sigma_{th}$ for the regions is $\sim$ 0.06 \kms and 0.05 \kms, respectively, and hence negligible, leading $\sigma_{nt}$ $\sim$ $\sigma_{obs}$.  
 This is an indication of the presence of turbulence in CC and NES, which has also been reported in \cite{Rawat_2024} for the C1 clump.

\begin{table*}
\caption{Parameters estimated for CC and NES.}
\begin{tabular}{|p{1.0cm} p{6cm} p{1.7cm} p{2.5cm} p{3.7cm}|} 
\hline
\hline

No & Parameter & Unit & CC & NES \\ 
\hline
1 & Effective radius \ \ \ \ ($R_{eff}$) & pc & 0.9 $\pm$ 0.1  & semi-minor axis ($r_1$) = 0.35 $\pm$ 0.03, semi-major axis ($r_2$) = 1.10 $\pm$ 0.09\\ 
2 & Mean column density \ \ \ \  ($N(H_2)$) & \cms  & (5.8 $\pm$ 2.2)$\times$ 10$^{21}$  & (7.1 $\pm$ 2.7) $\times$ 10$^{21}$\\
3 & Number density \ \ \ \  ($n_{H_2}$) & \cmq  & 1560 $\pm$ 780  & 3290 $\pm$ 1645\\
4 & Mass \ \ \ \ ($M$) & \Ms & 330 $\pm$ 148 & 188 $\pm$ 85\\
5 & Mean dust temperature \ \ \ \ ($T_d$) & K & 16.8 & 13.0\\
6 & Observed velocity dispersion \ \ \ \  ($\sigma_{obs}$) & \kms & 0.69 $\pm$ 0.05 & 0.41 $\pm$ 0.03\\
7 & Thermal velocity dispersion \ \ \ \ ($\sigma_{th}$) & \kms & 0.06 & 0.05\\
8 & Non-thermal velocity dispersion \ \ \ \  ($\sigma_{nt}$) & \kms & 0.69 $\pm$ 0.05 & 0.41 $\pm$ 0.03\\

\hline  

  &\multicolumn{3}{c}{Structure function analysis}   \\
\hline 

1 & Turbulent-to-ordered magnetic ﬁeld energy ratio \ \ \ \
 $\left(\frac{\braket{\delta B^2}^{1/2}}{B_0}\right)$ &   & 0.43 $\pm$ 0.01 & 0.46 $\pm$ 0.04\\
2 & Angular dispersion \ \ \ \ ($\sigma_{\theta}$) & degrees & 22.7 $\pm$ 0.6  & 23.9 $\pm$ 1.5\\
3 & Plane-of-sky magnetic ﬁeld strength \ \ \ \ ($B_{pos}$) & $\mathrm{\mu}$G & 24.0 $\pm$ 6.0 & 20.0 $\pm$ 5.0\\
4 & Mass-to-flux ratio \ \ \ \ ($\lambda_B$) & & 1.8 $\pm$ 0.8 & 2.7 $\pm$ 1.2\\
  
5 & \alf velocity \ \ \ \ ($V_A$) & \kms & 1.0 $\pm$ 0.4 & 0.6 $\pm$ 0.2\\
6 & \alf mach number \ \ \ \ ($\mathcal{M}_A$) & & 1.2 $\pm$ 0.4 & 1.2 $\pm$ 0.4\\ 
7 & Magnetic pressure \ \ \ \ ($P\rm{_B}$) & $\rm{dyne\, cm^{-2}}$ & (3.7 $\pm$ 1.9) $\times$ 10$^{-11}$ & (2.6 $\pm$ 1.3) $\times$ 10$^{-11}$\\
8 & Turbulent pressure \ \ \ \ ($P\rm{_{turb}}$) & $\rm{dyne\, cm^{-2}}$ & (5.2 $\pm$ 2.7) $\times$ 10$^{-11}$ & (2.5 $\pm$ 1.3) $\times$ 10$^{-11}$\\
\hline  

  &\multicolumn{3}{c}{Virial balance} \\
\hline
1 & Kinetic energy \ \ \ \ ($E_K$) & J & (4.7 $\pm$ 2.2) $\times$ 10$^{38}$ & (6.3 $\pm$ 3.0) $\times$ 10$^{37}$\\
2 & Magnetic energy \ \ \ \ ($E_B$) & J & (3.3 $\pm$ 3.0) $\times$ 10$^{38}$ & (7.0 $\pm$ 5.0) $\times$ 10$^{37}$\\
3 & Gravitational energy \ \ \ \ ($E_G$) & J & (10.4 $\pm$ 9.0) $\times$ 10$^{38}$ & (14 $\pm$ 12) $\times$ 10$^{37}$\\ 
4 & Kinetic virial parameter \ \ \ \ ($\alpha_{vir, k}$) & & 0.9 $\pm$ 0.4 & 0.9 $\pm$ 0.4\\
5 & Total virial parameter \ \ \ \ ($\alpha_{vir, tot}$) & & 1.2 $\pm$ 0.6 & 1.4 $\pm$ 0.7\\

\hline
\hline

\end{tabular}

\label{tab:results}

\end{table*}

\subsection{Magnetic field strength}
\label{Sect:mag}

\cite{Davis_1951} and \cite{Chand_1953} proposed a method to estimate the plane-of-sky component of the magnetic field ($B_{pos}$), known as the Davis-Chandrasekhar-Fermi (DCF) method, which is based on the assumption that the turbulence-induced \alf waves perturb the ordered B-field structure. Therefore, there will be a distorted component of the B-field that would appear as an irregular scatter in polarization angles in comparison to those that are produced by large-scale ordered B-field. Thus, the DCF method implies that the ratio of turbulent ($\delta B$) to ordered B-field ($B_0$) is proportional to the ratio of non-thermal velocity dispersion ($\sigma_{nt}$) to \alf velocity ($V_A = B_0/\sqrt{4 \pi \rho}$, $\rho$ is the gas mass density), i.e. $\frac{\delta B}{B_0} = \frac{\sigma_{nt}}{V_A}$. Also, the dispersion in the B-field position angles ($\sigma_\theta$) about the large-scale ordered B-field is assumed as $\sigma_\theta = \frac{\delta B}{B_0}$. Using these relations, the plane-of-sky magnetic field component, $B_{pos}$ can be estimated as

\begin{equation}
    B_{pos} = Q \sqrt{4 \pi \rho}\, \frac{\sigma_{nt}}{\sigma_\theta},
\end{equation} 
where $Q$ is the correction factor for the line-of-sight and beam-integration effects \citep{Ostriker_2001}. The studies show that the beam-integration effect can lead to an underestimation of angular dispersion in polarization angles, resulting in an overestimation of the magnetic field strength \citep{Ostriker_2001, Padoan_2001, Houde_2009}. 
To determine the angular dispersion, there are different statistical methods \citep[see][] {Hild_2009, Houde_2009, Pattle_2017, Liu_2022}, which are used in the literature. In this work, we use the structure-function \citep[SF;][]{Hild_2009} method, which gives the $\frac{\delta B}{B_0}$ ratio by accounting for the spatial variation of position angles.\\ 

\subsubsection{Structure function analysis}

In the SF method, the magnetic field is assumed to be composed of a large-scale structured ﬁeld and a turbulent ﬁeld that are statistically independent. The distinctive behaviour of the two components enables them to distinguish and extract the turbulent component, facilitating the computation of $\sigma_\theta$. 
The SF method computes the difference in position angles, $\Delta \phi(l) \equiv \phi(\bf x) - \phi(x + \textbf{\emph{l}})$, between the $N(l)$ pairs of pixels separated by $l$ = $|\textbf {\emph{l}}|$, using the following function:

\begin{equation}
    \braket{\Delta \phi^2(l)}^{1/2} \equiv \left(\frac{1}{N(l)} \sum_{i=1}^{N(l)} [\phi(\bf x) - \phi(x + l)]^2\right)^{1/2}.
\end{equation} 

\noindent This function is referred to as the "angular dispersion function". We want to point out that the polarization position angles are used here for the dispersion function. The angular dispersions, $\Delta \phi$, are kept $\leq$ 90\degree, to avoid the effect of the $\pm$ 180\degree~ambiguity of the magnetic field lines. Under the limit, $\delta < l << d$, the square of the angular dispersion function, known as the "structure function" is characterised by \citep{Hild_2009} 

\begin{equation}
    \braket{\Delta \phi^2(l)}_{tot} - \sigma^2_M(l) \simeq b^2 + m^2l^2, 
\label{eq:model}
\end{equation} where $\delta$ is the correlation length of the turbulent component, and $d$ is the typical length for variation in large-scale B-field. The quadratically added terms in the dispersion function, $m^2 l^2$ and $b^2$, are the contribution from the $B_0$ and $\delta B$, respectively. The $B_0$ is expected to increase almost linearly with slope $m$ for $l << d$, and $b$ is a constant turbulent contribution for $l > \delta$ \citep[for details, see][]{Hild_2009}. The $\sigma^2_M(l)$ is the contribution from the measured uncertainty in the position angles.\\

The turbulent to large-scale magnetic field strength is given by \citep{Hild_2009}

\begin{equation}
    \frac{\braket{\delta B^2}^{1/2}}{B_0} = \frac{b}{\sqrt{2-b^2}},
\label{eq:turb}
\end{equation}and $B_0$ can be estimated by using the modified DCF relation:

\begin{equation}
    B_0 \simeq \sqrt{(2-b^2)4\pi \mu m_H n_{H_2}}\,  \frac{\sigma_{nt}}{b}.
\label{eq:B0}
\end{equation} 
Using the $Q$ correction factor, we can determine the plane-of-sky magnetic field strength:

\begin{equation}
    B_{pos} = Q B_0,
\label{eq:Bpos}
\end{equation} where $Q$ is taken to be 0.5 \citep{Heitsch_2001, Ostriker_2001}.\\

\begin{figure*}
    \centering
    \includegraphics[width=8.5cm]{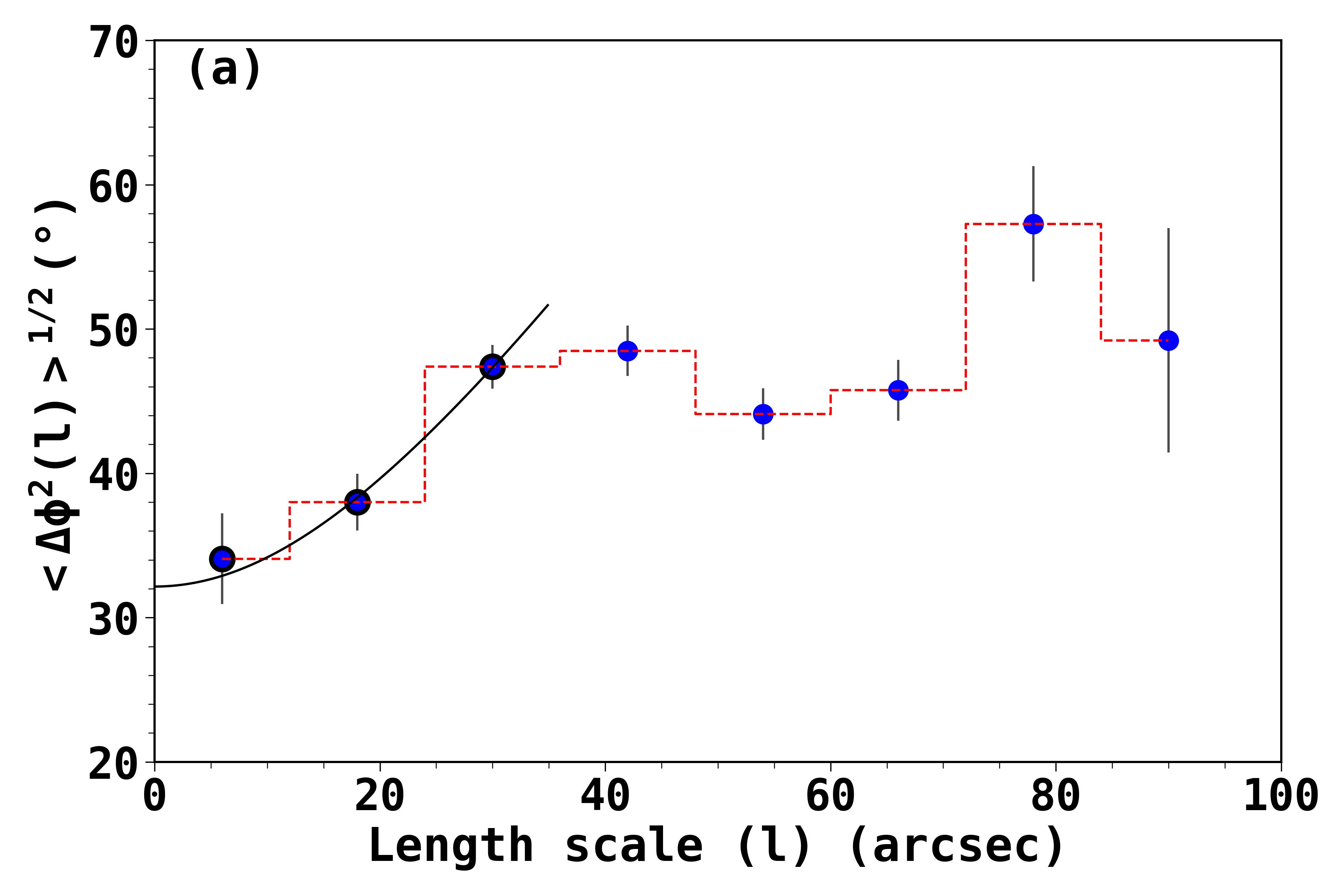}
    \includegraphics[width=8.5cm]{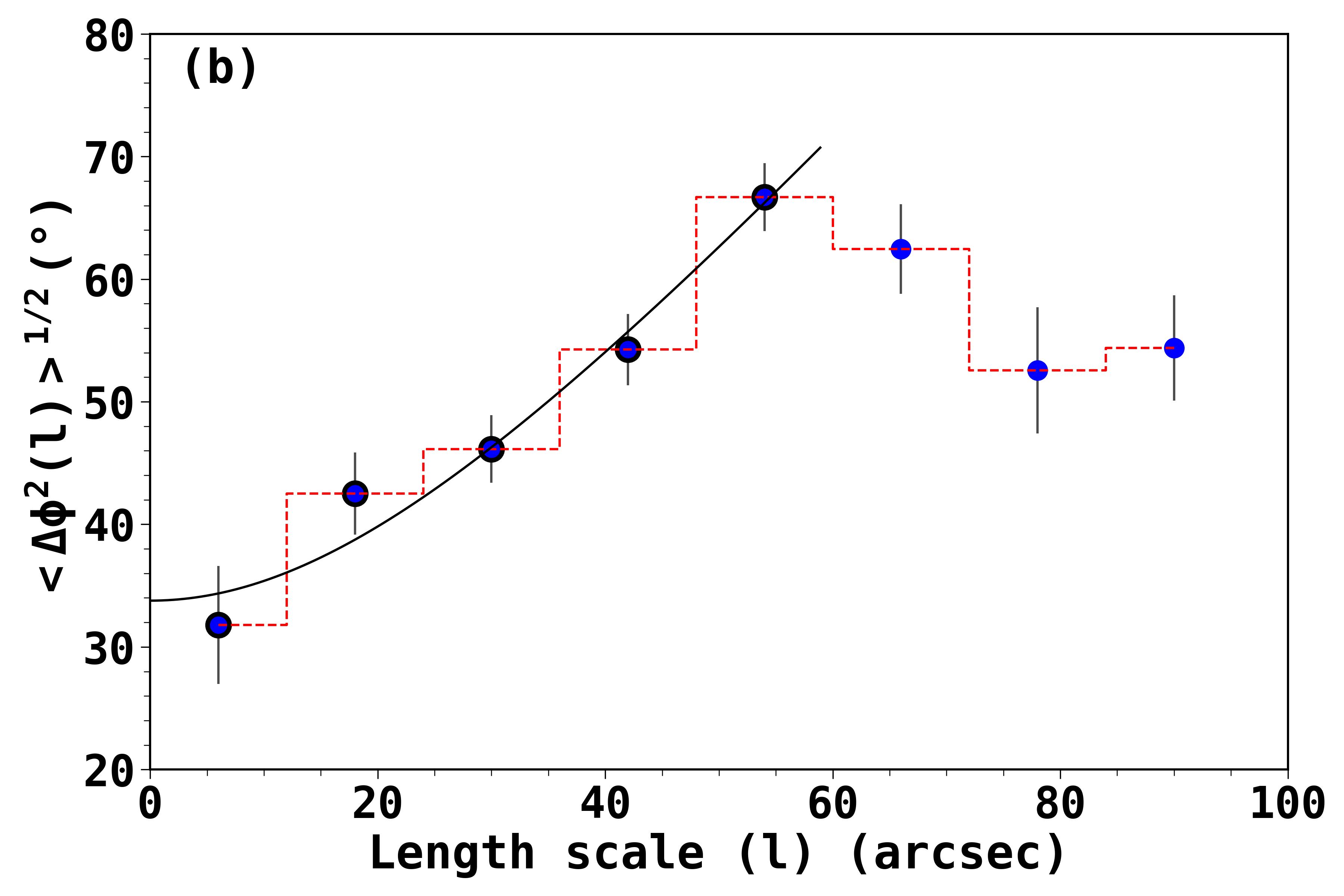}
    \caption{The angular dispersion function for (a) 
 CC  and (b) NES. The solid curve represents the best-fit model to the data, and the points used for fitting are shown in black encircled circles. The intercept of the best-fit model ($l$ = 0) gives the turbulent contribution to the total angular dispersion. The error bars denote the statistical uncertainties after binning and propagating the individual measurement uncertainties.}
    \label{fig:SF}
\end{figure*}

We calculated the dispersion function corrected by measurement uncertainty, i.e. $\braket{\Delta \phi^2(l)}_{tot} - \sigma^2_M(l)$ with a bin size of 12\arcsec. We used various bin sizes and found that the fit is converged, and fitting errors are the least for 12\arcsec. Fig. \ref{fig:SF} shows the dispersion in position angles as a function of the length scale for CC and NES. 
We fitted the dispersion function with the model defined in equation \ref{eq:model} using least square fit, over the first few data points to ensure the limit, $l << d$. The best-fits turbulent component, $b$, for CC and NES are 32\degree.1 $\pm$ 0\degree.9 and 33\degree.8 $\pm$ 2\degree.1, respectively. The dispersion in position angles can be obtained as $\sigma_\theta = b/\sqrt{2}$, which is around $\sim$22\degree.7 $\pm$ 0\degree.6 and 23\degree.9 $\pm$ 1\degree.5 for CC and NES, respectively. These values are close to the maximum value at which the DCF methods would give reliable results \citep[$\sigma_{\theta} \leq$ 25\degree,][]{Ostriker_2001} if a correction factor of 0.5 is applied. Using equation \ref{eq:turb}, \ref{eq:B0}, and \ref{eq:Bpos}, we calculate the $\frac{\delta B}{B_0}$ ratio to be around $\sim$0.43 $\pm$ 0.01 and 0.46 $\pm$ 0.04, and $B_{pos}$ to be around $\sim$24.0 $\pm$ 6.0 \muG~and 20.0 $\pm$ 5.0 \muG, for CC  and NES, respectively. All the estimated parameters are given in Table \ref{tab:results}. 

\section{Discussion}
\label{discussion}
The plane-of-sky magnetic field strength for the CC and NES regions is around $\sim$24 \muG~and 20 \muG, respectively. Given the uncertainty of at least a factor of 2 associated with the B-field estimation by the DCF method \citep{Crutcher_2012},  the estimated B-field strengths are within the range of $\sim$10$-$100 \muG~observed in star-forming regions \citep[][]{chapman_2011, Crutcher_2012, Pattle_2022}. The values are also consistent (i.e. within a factor of 1.5) with the upper limits of the B-field values from \cite{Crutcher_2010} relation for the respective density of the regions.  

\subsection{Correlation between magnetic fields, intensity gradients, and local gravity}
\label{U_shape}

\begin{figure}
    \centering
    \includegraphics[width=8.5cm]{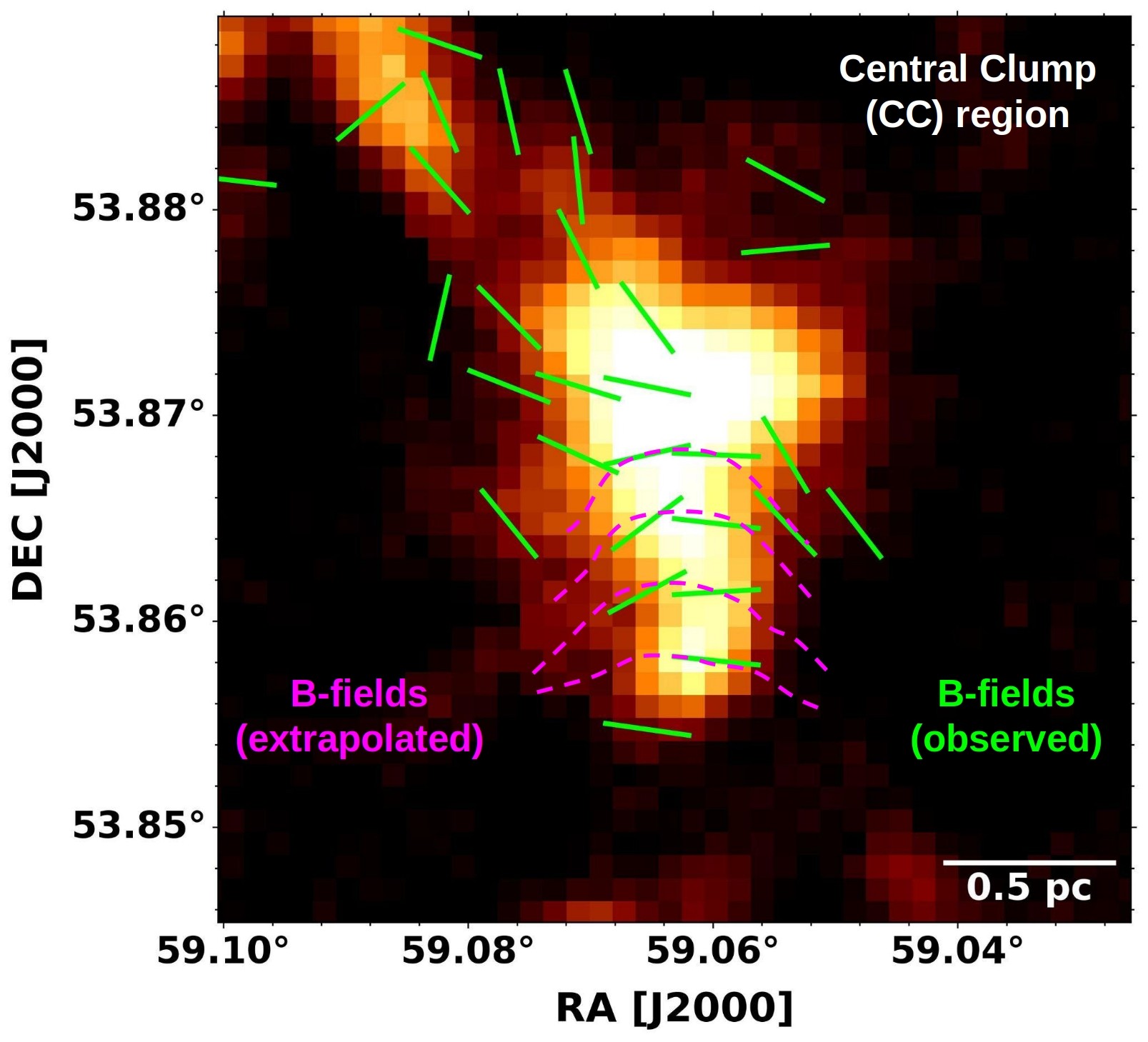}
    \caption{Stokes I map of the Central clump region of \cloud, over which the observed B-field segments are shown (green segments). The B-field segments show a ``U" shaped geometry, sketched with observed segments and shown by magenta dashed curves, which may be caused by the drag of gravitational converging flows towards the centre of the cloud.}
    \label{fig:ushape}
\end{figure}

Due to the low statistics in the polarization data, it is difficult to conclusively comment on the overall morphology of the B-field, intensity gradients, and local gravity. However, through this comparison, some inferences can be drawn over their relative spatial variance.
Fig. \ref{fig:B_vs_int} and \ref{fig:gravity_map} shows a correlation between the B-field, intensity gradients, and local gravity in the CC and WC regions, whereas the differences in their position angles are relatively higher in the NES region. This correlation can be due to the collapse of the clumps where gravity has pulled in and aligned B-field lines with the intensity gradients, in the CC and WC regions \citep{Koch_2013, Wang_2019}. However, in the outer diffused regions, like the elongated structures, the field lines are not yet that much affected by the local gravity.\\ 

Some simulations of the global collapse of magnetized clouds have found that the gravitational flows from large scale to small scale, i.e. from filaments to clumps/cores, can drag the magnetic field lines along the flow, causing a ``U" shaped geometry of the field lines across the filament spine \citep{Gomez_2018, sema2019}. We also found evidence of ``U" shaped geometry of B-field lines, at the bottom of the elongated part of the central clump. Fig. \ref{fig:ushape} shows the zoomed-in view of the central clump region, in which the ``U" shaped geometry is sketched from the observed B-field orientations and is shown by magenta curves. High-resolution and sensitivity observations would be required to ascertain the observed morphology. A similar effect of gravity over the B-field morphology has also been found in other observational works \citep{Tang_2019, Wang_2020a, Wang_2020b, Beuther_2020, Busq_2020, Pillai_2020}.\\

In \cloud, \cite{Rawat_2024}, using MWISP CO data, identified filamentary gas flows exhibiting noticeable velocity gradients as they move towards the hub. They found that the velocity gradient increases towards the hub, with a measured value of $\sim$0.2 \kms pc$^{-1}$ in the proximity of the hub. The aforementioned findings give evidence of accreting gas flows along the filaments towards the cloud's centre, which can affect the B-field morphology by dragging them along the flow. 
Future high-resolution dust continuum and polarimetric observations might be able to reveal a significant number of polarization vectors at the clump/core scale to better resolve the substructures and the B-field morphology around the hub of \cloud.


\subsection{Gravitational stability}
\label{comp}
\subsubsection{Mass-to-flux ratio}

In order to investigate whether the magnetic field can provide stability to the regions against gravitational collapse, we determine the mass-to-flux ratio \citep{Crutcher_2004}. It is generally calculated as a dimensionless critical stability parameter, $\lambda_B$, which is basically the comparison of the mass to magnetic flux ratio with the critical ratio \citep{Crutcher_2004}:

\begin{equation}
\label{eq:lambda}
    \lambda_B = \frac{(M/\phi)_{obs}}{(M/\phi)_{crit}} = 7.6 \times 10^{-21} \ \ \frac{N(H_2) (\cms)}{B_{pos} (\mu G)},
\end{equation} where $N(H_2)$ is the mean column density of the region. A clump or dense core is magnetically supercritical ($\lambda_B$ $>$ 1) if the magnetic field is not strong enough to support the system against gravitational collapse, whereas a strong magnetic field would make the system magnetically subcritical ($\lambda_B$ $<$ 1), i.e. stable against collapse. Using the mean $N(H_2)$ $\sim$(5.8 $\pm$ 2.2) $\times$ 10$^{21}$ \cms and (7.1 $\pm$ 2.7) $\times$ 10$^{21}$ \cms, we calculated the critical parameter to be around $\sim$1.8 $\pm$ 0.8 and 2.7 $\pm$ 1.2 for CC  and NES, respectively. The critical parameter shows that both regions are magnetically supercritical. In general, a statistical correction factor is applied to $\lambda_B$ to account for bias due to geometric effects \citep{Crutcher_2004}. Different correction factors are suggested for different geometry of clumps with respect to magnetic field, e.g., $\pi/4$ for spherical clump and $1/3$ for oblate spheroid with major-axis perpendicular to the mean B-field \citep{Crutcher_2004}, and $3/4$ for prolate spheroid with major-axis parallel to the mean B-field \citep{Planck_2016}. Using these correction factors, the mass-to-flux ratios become subcritical to transcritical/supercritical in the range of 0.6 to 1.4 for CC with mean $\sim$ 1.1, and 0.9 to 2.1 for NES with mean $\sim$ 1.7. However, if we consider the overestimation of $B_{pos}$ in the DCF method itself, our estimated mass-to-flux ratios could also be a lower limit. 
We want to acknowledge here that the estimated B-field strengths and mass-to-flux ratios are only averaged values over the selected regions. 
The regions can still be super-critical inside but sub-critical in the outer part, as also shown in a recent simulation by \cite{Gomez_2021}.    



\subsubsection{Turbulence versus magnetic field}
\label{turb_vs_B}

Simulations suggest that turbulence plays a dual role in the clouds and their substructures by providing turbulent support against the gravitational collapse at a large scale while producing compressions and shocks at small scales, that create density enhancements and trigger the star formation process \citep{Maclow_klessen_2004, Balles_2007, Henne_2012, klessen_2016}. Also, whether it is turbulence (weak B-field models) or magnetic field (strong B-field models) or both, is a subject of investigation regarding their respective roles in the formation of clumps, cores, and subsequent star formation within these structures. In order to investigate their impact, we need to find out the relative strength of turbulence in comparison to the magnetic field.\\

The \alf Mach number (\mach) infers the relative importance of turbulence and magnetic field in molecular clouds and is defined as \mach = $\sqrt{3} \sigma_{nt}/V_A$. The \alf velocity is calculated as, $V_A = B_{tot}/\sqrt{4\pi\rho}$. The total B-field strength ($B_{tot}$) of the regions can be determined by using a statistical relation, $B_{tot} = (4/\pi) B_{pos}$ \citep{Crutcher_2004}. For CC and NES, the $B_{tot}$ is found to be around 31 $\pm$ 8~\muG~and 26 $\pm$ 6~\muG, respectively. Using the mass density estimated within the dimensions of two regions (given in Table \ref{tab:results}) and corresponding $B_{tot}$ values, the $V_A$ for CC  and NES are found to be $\sim$1.0 $\pm$ 0.4 \kms and 0.6 $\pm$ 0.2 \kms, respectively. Then, using the corresponding $\sigma_{nt}$ values, the $\mach$ is calculated to be around $\sim$1.2 $\pm$ 0.4 for both the two regions. A star-forming region is super-Alfv$\acute{\text{e}}$nic if \mach $>1$, which means that the turbulent pressure is higher than the magnetic pressure. Conversely, it will be sub-Alfv$\acute{\text{e}}$nic if \mach $<1$, which means that the magnetic pressure is higher than the turbulent pressure. In the present case, both the regions are trans Alfv$\acute{\text{e}}$nic.\\

We also calculated the magnetic and turbulent pressures using the relations:
\begin{equation}
  P_B = \frac{B_{tot}^2}{8\pi} \ \  \ \ \mathrm{and} \ \ \ \ P_{t} = \frac{3}{2}\rho \sigma_{nt}^2\, \rm{(Spherical)},    
\end{equation} 

\begin{equation*}
  \qquad \qquad \qquad \qquad \quad \; = \rho \sigma_{nt}^2\ \ \ \ \rm{(Cylindrical)},   
\end{equation*} 
here, a factor of 3/2 is included to estimate the total turbulent pressure by assuming the non-thermal velocity dispersion to be isotropic in spherical geometry. For the CC region, the magnetic and turbulent pressure is found to be $\sim$(3.7 $\pm$ 1.9) $\times$ 10$^{-11}$ dyne \cms and $\sim$(5.2 $\pm$ 2.7) $\times$ 10$^{-11}$ dyne $\rm{cm^{-2}}$, respectively. For the NES region, $P_B$ and $P_{t}$ are $\sim$(2.6 $\pm$ 1.3) $\times$ 10$^{-11}$ dyne \cms and (2.5 $\pm$ 1.3) $\times$ 10$^{-11}$ dyne $\rm{cm^{-2}}$, respectively. In the CC region, the turbulent pressure is higher than the magnetic pressure by a factor of $\sim$1.4, while in NES, the pressure values are similar. Also, the thermal pressure ($\sim$$\rho \sigma_{th}^2$) in both regions, is much smaller than the turbulent and magnetic pressure, which shows that the thermal energy plays a negligible role in energy balance. So, from the \alf Mach number and pressure estimation, it seems that the turbulence is slightly more dominant in comparison to the magnetic field in CC, i.e. in the centremost part of \cloud, while it is similar in the NES.   

\subsubsection{Virial analysis}
\label{Virial}
The virial theorem is a principle that relates the average kinetic energy ($E_K$) and magnetic field energy ($E_B$) of a system to its average gravitational potential energy ($E_G$), which provides insights into the stability and energy distribution of the system. The Virial theorem is written as:

\begin{equation}
    \frac{1}{2}\frac{d^2\mathcal{I}}{dt^2} = 2E_K + E_B + E_G,
\end{equation} where $\mathcal{I}$ is the moment of inertia. The surface energy terms here are neglected. The kinetic energy term is given by \citep{Fiege_2000}

\begin{equation}
    E^{sph}_{K} = \frac{3}{2} M \sigma_{obs}^2\ \ \ \  \rm{(Spherical)},
\end{equation}

\begin{equation*}
    E^{cyl}_{K} = M \sigma_{obs}^2\ \ \ \ \ \ \ \ \rm{(Cylindrical)},
\end{equation*}
where $M$ is the mass and $\sigma_{obs}^2 = \sigma_{th}^2 + \sigma_{nt}^2$. The magnetic field energy is given by

\begin{equation}
    E_B = \frac{1}{2} M V_{A}^2
\end{equation}
and the gravitational potential energy is given by

\begin{equation}
    E^{sph}_G = -\frac{(3-a)}{(5-2a)} \frac{G M^2}{R}\ \ \ \  \rm{(Spherical)},
\end{equation}

\begin{equation*}
    E^{cyl}_{G}\; = -\frac{G M^2}{L}\qquad \qquad \ \  \rm{(Cylindrical)},
\end{equation*}
where $G$ is the gravitational constant, $R$, is the effective radius of the sphere, and $L$ is the length of the cylinder. Here $a$ is the density profile index of the sphere ($\rho \propto r^{-a}$). 
The energy values for the two regions are listed in Table \ref{tab:results}. We found that for CC, $|E_G|$ $>$ $E_K$ $>$ $E_B$ (i.e. 10.4:4.7:3.3), while for NES, $|E_G|$ $>$ $E_K$ $\simeq$ $E_B$ (i.e. 14:6.3:7.0), which restates the results of \alf Mach number and mass-to-flux ratio calculation, i.e. turbulence is slightly more dominant than the magnetic field in CC, while it is similar in NES, but overall the gravity is the dominant factor in both the regions. Nevertheless, the calculation of individual energy terms is an important exercise, enabling the direct comparison of various forces that govern the evolution of clumps/structures, expressed in the same units.\\

For a non-magnetized system ($E_B$ = 0), the stability criteria is given by 2$E_K$ $+$ $E_G$ $<$ 0, and based on this condition, the kinetic virial parameter is defined as \citep{Bert_1992}

\begin{equation}
    \alpha_{vir, k} = \frac{2E_K}{|E_G|} \ \  =  \frac{3(5-2a)}{(3-a)} \frac{R \sigma_{obs}^2} {GM}\; \rm{(Spherical)},
\label{eq:alpha_vir}
\end{equation}

\begin{equation*}
    \qquad \qquad \qquad \ \ \;  = \ \ \frac{2L}{GM} \sigma_{obs}^2\qquad \ \ \ \ \rm{(Cylindrical)}.
\end{equation*}
Using the $M$, $R$, $L$, and $\sigma_{obs}$ values of the regions (given in Table \ref{tab:results}) in equation \ref{eq:alpha_vir}, and adopting $a$ = 2 for spherical case, we calculate the $\alpha_{vir, k}$ to be around 0.9 $\pm$ 0.4 and 0.9 $\pm$ 0.4, respectively for CC and NES. The derived $\alpha_{vir, k}$ $<$ 2 means that in the case of a non-magnetized sphere ($E_B$ = 0), the thermal plus turbulent contribution is not enough to provide stability to the regions against the gravitational collapse \citep{Kauff_2013, Mao_2020}. Including the magnetic energy term in the stability criteria, i.e. 2$E_K$ $+$ $E_B$ $+$ $E_G$ $<$ 0, the total virial parameter is calculated using the modified relation \citep{Bert_1992, Pillai_2011, Sanhueza_2017}.

\begin{multline}
   \alpha_{vir_{tot}} = \frac{2E_K + E_B}{|E_G|}  \ \ =  \ \ \frac{3(5-2a)}{(3-a)} \frac{R}{GM} \left(\sigma_{obs}^2 + \frac{V_A^2}{6}\right)\\ \qquad \qquad \qquad \qquad \qquad \qquad \qquad \rm{(Spherical)},
\end{multline}

\begin{equation*}
   \qquad \qquad \qquad \qquad \quad \,  =  \ \ \frac{2L}{GM} \left(\sigma_{obs}^2 + \frac{V^2_A}{4}\right) \quad  \rm{(Cylindrical)}.
\end{equation*}
With the magnetic support, the $\alpha_{vir, tot}$ value is estimated to be around $\sim$1.2 $\pm$ 0.6 and 1.4 $\pm$ 0.7 for CC and NES, respectively. The $\alpha_{vir, tot}$ values for the regions are $<$ 2, which shows that the two regions are bound by gravity and thus can collapse to form stars. The virial analysis shows that the total kinetic energy ($E_{K}$, i.e. thermal plus turbulent) in both regions is not sufficient to support them against the gravitational collapse. While magnetic energy, combined with kinetic energy, is found to be comparable to gravitational potential energy.\\


In the present work, we have estimated magnetic field strengths and derived various parameters for the CC and NES regions. However, we want to stress that these results must be taken with caution as the measurements are uncertain within a factor two due to the inherent large uncertainty in the mass and density of the studied regions. Moreover, the modified DCF methods can be uncertain up to a factor of two or more \citep{Crutcher_2012}, and also, the B-field strength in the studied regions can be biased due to the limited number of B-field segments traced by our observations.  Due to all these uncertainties, the mass-to-flux ratio and virial status of the regions should be considered as qualitative indicators of the stability of the region. 
In the present case, we have used the generally accepted Q value of 0.5 \citep{Crutcher_2012} in our estimations, however, if we use Q = 0.4, suggested for parsec scale clumps \citep{Padoan_2001}, similar to our studied regions, the B-field strengths will reduce by a factor of $\sim$1.2. As a consequence, the CC and NES regions would become more magnetically supercritical. Future high-resolution and more sensitive observations would better constrain the magnetic field and turbulence properties of the hub. However, taking the measured mass-to-flux ratio and virial parameters at face value, it can be argued that, at present, gravity has overall an upper hand over magnetic and kinetic energies in CC and NES, which is consistent with the formation of a young cluster noticed by \citet{Rawat_2023} in the hub.

\subsection{The criticality of magnetic fields in hub-filamentary clumps}

Like the CC of \cloud, the clumps located at the junction of the hub-filamentary systems are known to be potential sites of cluster formation \citep[e.g.][]{Kum_2020}, because such clumps are attached to converging filamentary structures that fuel them with cold gaseous matter \citep[e.g.][]{Myers_2009, Li_2014, Tre_2019, Kum_2022}. Although physical processes that govern the star formation are scale-dependent, it is worthwhile to compare the global properties of the parsec or sub-parsec scale hub filamentary clumps studied with similar resolution.  In the literature, a few such clumps have been studied with JCMT/POL2, these are: IC 5146 E-hub \citep{Wang_2019}, G33.92$+$0.11 \citep{Wang_2020b}, Mon R2 \citep{Hwang_2022}, IC 5146 W-hub \citep{Chung_2022}, and SDC13 \citep{Wang_2022}. In the majority of these hub filamentary clumps, except Mon R2, the mass-to-flux ratio and/or virial analysis suggest the dominance or edge of gravitational energy over the magnetic and kinetic energies, similar to the CC region of \cloud. 
We found that all the aforementioned clumps are associated with either protostars or an embedded cluster \citep[e.g.][]{har_2008,Gut_2009,Per_2014,liu_2015}. Thus, it seems that, at least for those parsec scale hubs that are in the early stages of cluster formation, like the CC of \cloud, gravity has an upper hand on the energy budget of the system.  A large sample of hub-filamentary clumps of various evolutionary stages (e.g. from pre-stellar clumps to clumps hosting emerging clusters of different ages) would be valuable to study the time evolution of various physical processes that govern star formation and its evolution.

\section{Summary}
\label{summary}
We have performed the dust polarization observation of the central part of the \cloud~cloud to investigate the B-field morphology and its strength relative to gravity and turbulence, using JCMT SCUBA-2/POL-2 at 850 \mum. We specifically focused on the C1 clump, located at the hub of \cloud. The main results are summarised below:

\begin{itemize}
    \item The 850 \mum Stokes I intensity map reveals the presence of a central clump (CC), northeastern elongated structure (NES), and western clump (WC) around the hub of \cloud. 
    The B-field segments of CC and NES regions show mixed morphology, while the WC region shows converging B-field segments, mostly aligned along the southeast direction.
    

\item We found evidence of the depolarization effect, and from the Bayesian analysis over the non-debiased polarization data, found a power-law index, $\alpha$ = 0.6. Although this shows a decreasing level of dust grain alignment, but they can still be aligned with the magnetic field in the central high-density region of the cloud.  

    \item We compared the relative orientations of B-fields ($\theta_B$), intensity gradients ($\theta_{IG}$), and local gravity ($\theta_{LG}$) over the full map. In the CC and WC regions, the three factors are mostly correlated, while the difference in orientations is higher in the NES region. This suggests that gravity is dragging the intensity gradients and aligning them with the B-fields in the CC and WC clump, while the effect of gravity in NES is comparatively less significant.

    \item We constructed the $\Sigma_B$ map to see the localised B-field strength in comparison to local gravity, and found that for most of the parts, $\Sigma_B$ $<$ 1, i.e. gravitational force is dominant over the magnetic field force.

    \item We used the structure function analysis to determine the B-field strength, and as a result, found $B_{pos}$ for CC and NES to be around  $\sim$24.0 $\pm$ 6.0 \muG~and 20.0 $\pm$ 5.0 \muG, respectively. In both regions, the turbulent component relative to the organised magnetic field structure was determined to be approximately 0.4$-$0.5.

    \item The mass-to-flux ratio and \alf Mach number calculation for CC and NES shows that both regions are magnetically transcritical/supercritical and trans-Alfv$\acute{\text{e}}$nic. The turbulent pressure was found to be higher than the magnetic pressure in CC, while they are similar in NES. 

    \item The virial analysis shows that for CC, the $|E_G|$ $>$ $E_K$ $>$ $E_B$ (i.e. 10.4:4.7:3.3), while for NES, $|E_G|$ $>$ $E_K$ $\simeq$ $E_B$ (i.e. 14:6.3:7.0). The magnetic field and turbulence individually are not strong enough to provide stability to the regions against gravity.   Both regions were found to be bound by gravity. \\
      
   Overall, we find that currently, gravitational energy has an edge over the other energy terms of the hub region of \cloud, thereby continue to facilitate the growth of the young cluster in the hub, although we acknowledge that given the large uncertainties associated with our estimates, a conclusive answer would require further precise measurements of magnetic field and cloud properties.

\end{itemize}

\section*{ACKNOWLEDGEMENT}

We thank the anonymous referee for the comments and suggestions
that helped to improve the paper. The research work at the Physical Research Laboratory is funded by the Department of Space, Government of India. CE acknowledges the
financial support from grant RJF/2020/000071 as a part of the Ramanujan Fellowship awarded by the Science and Engineering Research Board (SERB), Department of Science and Technology (DST), Government of India. The James Clerk Maxwell Telescope is operated
by the East Asian Observatory on behalf of The National
Astronomical Observatory of Japan; Academia Sinica Institute
of Astronomy and Astrophysics; the Korea Astronomy and
Space Science Institute; the National Astronomical Research
Institute of Thailand; Center for Astronomical Mega-Science
(as well as the National Key R\&D Program of China with
No. 2017YFA0402700). Additional funding support is
provided by the Science and Technology Facilities Council of
the United Kingdom and participating universities and
organizations in the United Kingdom, Canada, and Ireland.
The authors wish to recognize and acknowledge the very
significant cultural role and reverence that the summit of Maunakea has always had within the indigenous Hawaiian
community. The data taken in this
paper was observed under the project code M22BP055. D.K.O. acknowledges the support of the Department of Atomic Energy, Government of India, under project identification No. RTI 4002. We thank Junhao Liu for the discussion on 
the DCF methods. This research made use of the data from the Milky Way Imaging Scroll Painting (MWISP) project, which is a northern galactic plane CO survey with the PMO-13.7m telescope. We are grateful to all the members of the MWISP working group, particularly the staff members at PMO-13.7m telescope, for their long-term support. MWISP was sponsored by the National Key R\&D Program of China with grant 2017YFA0402701 and CAS Key Research Program of Frontier Sciences with grant QYZDJ-SSW-SLH047.\\

\emph{Facilities}: JCMT, PMO

\section*{Data Availability}

The JCMT data used in this work can be shared on reasonable request. We also used the CO molecular data from PMO, which can be shared by the PMO database on reasonable request to the project team.



\bibliographystyle{mnras}
\bibliography{mybibliography.bib} 



\appendix

\section{Intensity gradients versus Local gravitational field}
\begin{figure*}
    \centering
    \includegraphics[width=8.0cm]{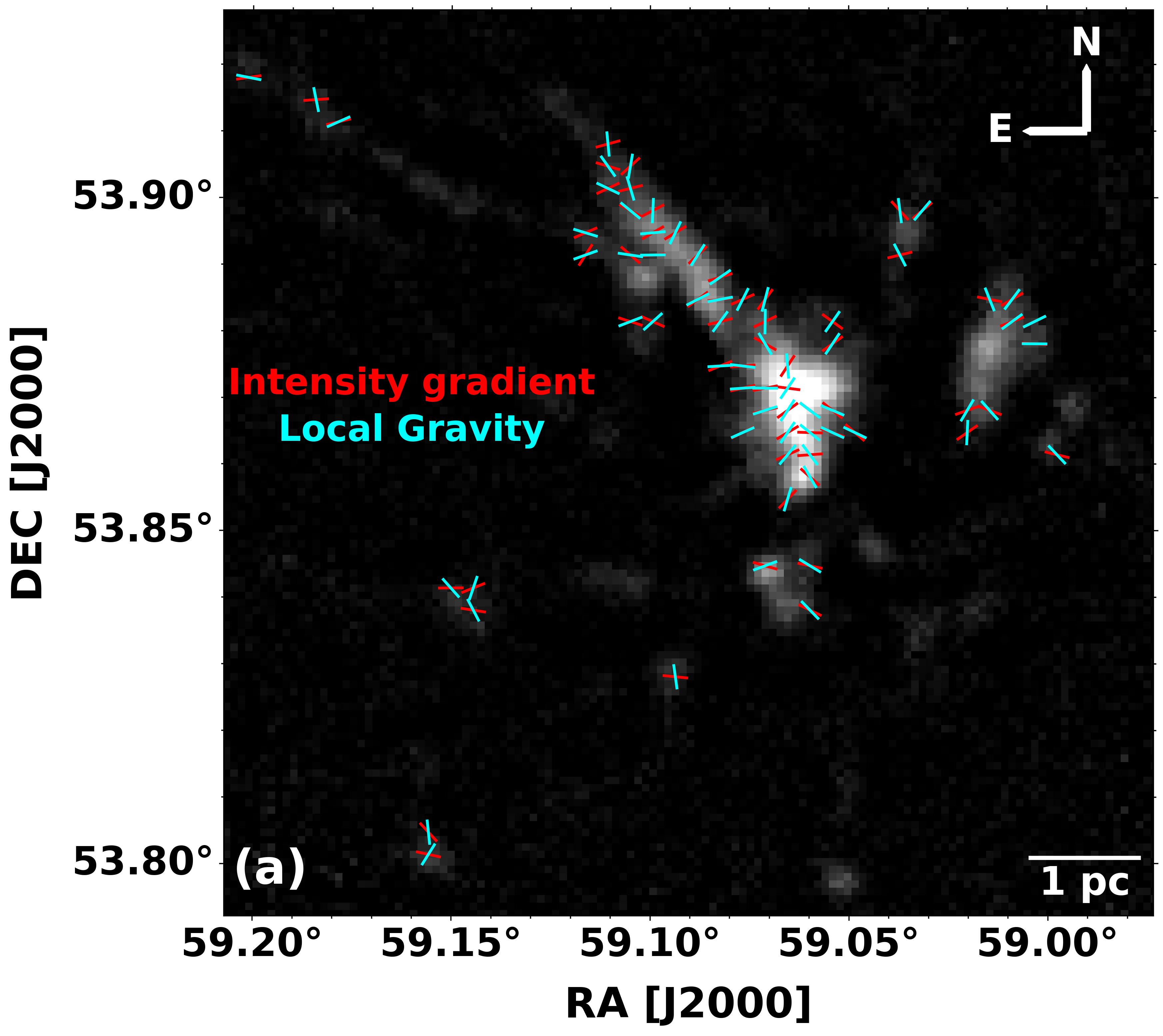}
    \includegraphics[width=8.5cm]{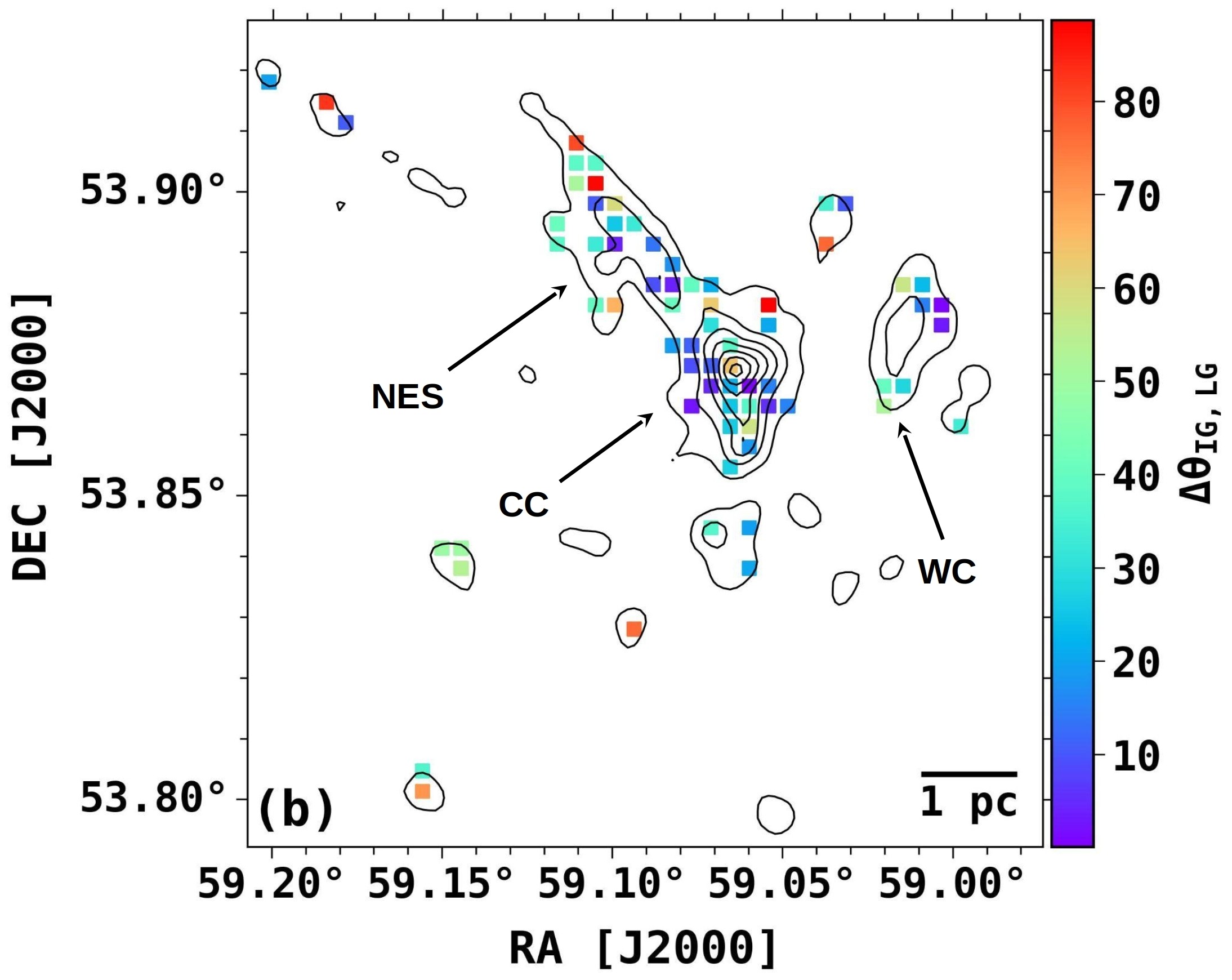}
    \caption{(a) The orientations of the intensity gradients (red segments) and local gravity (cyan vectors) are overlaid on the 850 \mum Stokes I map. (b) The distribution of the offset between the position angles of the intensity gradients and local gravity, i.e., $\Delta \theta_{IG, LG}$ = $|(\theta_{IG} - \theta_{LG})|$ over the contours of 850 \mum Stokes I map. The contour levels are same as in Fig. \ref{fig:polmap}. }
    \label{fig:IG_vs_LG}
\end{figure*}

Fig. \ref{fig:IG_vs_LG}a shows the relative orientations of intensity gradient and local gravity, and Fig. \ref{fig:IG_vs_LG}b shows the distribution of the angular difference between their orientations, i.e. $\Delta \theta_{IG, LG}$. Similar to $\Delta \theta_{B, IG}$ and $\Delta \theta_{B, LG}$, the correlation between the angles ($\theta_{IG}$ and \textbf{$\theta_{LG}$}) is better in the CC and WC regions in comparison to NES region.

\bsp	
\label{lastpage}

\end{document}